\documentclass[prd,preprint,floatfix,nofootinbib,superscriptaddress,showkeywords]{revtex4} 

\pdfoutput=1
\usepackage[caption=false]{subfig}
\usepackage{amsmath}
\usepackage{amsfonts}
\usepackage{amssymb}
\usepackage{float}
\usepackage{color}
\usepackage{graphicx}
\usepackage{graphics}
\usepackage{hyperref}
\usepackage{adjustbox,array}
\usepackage[utf8x]{inputenc} 
\usepackage{epstopdf}
\usepackage{multirow}
\usepackage{slashed}
\usepackage{booktabs} 
\usepackage{cancel}
\usepackage{mathrsfs}
\usepackage{csquotes}

\definecolor{rosso}{cmyk}{0,1,1,0.4}
\definecolor{rossos}{cmyk}{0,1,1,0.55}
\definecolor{rossoc}{cmyk}{0,1,1,0.2}
\definecolor{blu}{cmyk}{1,1,0,0.3}
\definecolor{magebtas}{cmyk}{0.1,1,0.1,0.2}
\definecolor{bluc}{cmyk}{1,1,0,0.1}
\definecolor{verde}{cmyk}{0.92,0,0.59,0.25}
\definecolor{verdec}{cmyk}{0.92,0,0.59,0.15}
\definecolor{verdes}{cmyk}{0.92,0,0.59,0.4}

\AtBeginDocument{
\heavyrulewidth=.08em
\lightrulewidth=.05em
\cmidrulewidth=.03em
\belowrulesep=.65ex
\belowbottomsep=0pt
\aboverulesep=.4ex
\abovetopsep=0pt
\cmidrulesep=\doublerulesep
\cmidrulekern=.5em
\defaultaddspace=.5em
}

\hypersetup{
 colorlinks=true,
 linkcolor=verdec,
 urlcolor=verdec,
 citecolor=verdec
}

\newcommand{\eq}[1]{Eq.~(\ref{#1})}
\newcommand{\fig}[1]{Fig.~\ref{#1}}

\newcommand{\gsim}{\gtrsim}
\newcommand{\lsim}{\lesssim}

\newcommand{\lf}{\left(}
\newcommand{\ri}{\right)}

\newcommand{\nn}{\nonumber}

\newcommand{\sqt}{\sqrt{2}}

\newcommand{\rr}{{\gamma\gamma}}

\renewcommand{\lg}{\mathscr{L}} 

\newcommand{\mco}{\mathcal{O}}

\newcommand{\br}{{\rm Br}}

\newcommand{\hc}{{\rm H.c.}}
\newcommand{\tp}{{\rm type}}

\newcommand{\fb}{{\;{\rm fb}}}
\newcommand{\ab}{{\;{\rm ab}}}

\newcommand{\iab}{{\;{\rm ab}^{-1}}}

\newcommand{\gev}{{\;{\rm GeV}}}
\newcommand{\tev}{{\;{\rm TeV}}}

\newcommand{\beq}{\begin{equation}}
\newcommand{\eeq}{\end{equation}}
\newcommand{\bea}{\begin{eqnarray}}
\newcommand{\eea}{\end{eqnarray}}
\newcommand{\barr}{\begin{array}}
\newcommand{\earr}{\end{array}}
\newcommand{\bc}{\begin{center}}
\newcommand{\ec}{\end{center}}
\newcommand{\bit}{\begin{itemize}}
\newcommand{\eit}{\end{itemize}}
\newcommand{\ben}{\begin{enumerate}}
\newcommand{\een}{\end{enumerate}}

\newcommand{\al}{\alpha}
\newcommand{\bt}{\beta}

\newcommand{\dt}{\delta}

\newcommand{\sg}{\sigma}

\newcommand{\kp}{\kappa}

\newcommand{\gm}{\gamma}

\newcommand{\lm}{\lambda}
\newcommand{\lmh}{\hat{\lambda}}

\newcommand{\lmc}{\Lambda_{\rm cut}}

\newcommand{\met}{E_T^{\rm miss}}

\newcommand{\hsm}{{h_{\rm SM}}}

\newcommand{\ch}{H^\pm}

\newcommand{\wmp}{W^\mp}
\newcommand{\mh}{m_{h}}
\newcommand{\mhsm}{m_{125}}
\newcommand{\mch}{M_{H^\pm}}
\newcommand{\mhh}{M_{H}}
\newcommand{\ma}{M_{A}}

\newcommand{\tb}{t_\beta}

\newcommand{\cb}{c_\beta}
\renewcommand{\sb}{s_\beta}

\newcommand{\cba}{c_{\beta-\alpha}}
\newcommand{\sba}{s_{\beta-\alpha}}

\newcommand{\mmu}      {{\mu^+ \mu^-}}

\newcommand{\ttau}      {{\tau^+\tau^-}} 

\newcommand{\ttop}      {{t\bar{t}}}

\newcommand{\bb}      {{b \bar{b}}}

\newcommand{\qq}      {{q \bar{q}}}

\definecolor{mint}{rgb}{0.24, 0.71, 0.54}

\begin{document}

\title{\color{verdes} Disentangling the high- and low-cutoff scales\\
via the trilinear Higgs couplings\\
in the type I two-Higgs-doublet model}
\author{Sin Kyu Kang}
\email{skkang@snut.ac.kr}
\address{School of Natural Science, Seoul National University of Sciene and Technology, Seoul 139-743, Korea}
\author{Jinheung Kim}
\email{jinheung.kim1216@gmail.com}
\address{Department of Physics, Konkuk University, Seoul 05029, Republic of Korea}
\author{Soojin Lee}
\email{soojinlee957@gmail.com}
\address{Department of Physics, Konkuk University, Seoul 05029, Republic of Korea}
\author{Jeonghyeon Song}
\email{jhsong@konkuk.ac.kr}
\address{Department of Physics, Konkuk University, Seoul 05029, Republic of Korea}
\begin{abstract}
The type I two-Higgs-doublet model in the inverted Higgs scenario can retain the theoretical stability all the way up to the Planck scale. The Planck-scale cutoff, $\Lambda_{\rm cut}^{\rm Planck}$, directly impacts the mass spectra such that all the extra Higgs boson masses should be light below about 160 GeV. However, the observation of the light masses of new Higgs bosons does not indicate the high-cutoff scale because a low-cutoff scale can also accommodate the light masses. Over the viable parameter points that satisfy the theoretical requirements and the experimental constraints, we show that the trilinear Higgs couplings for low $\Lambda_{\rm cut}$ are entirely different from those for the Planck-scale cutoff. The most sensitive coupling to the cutoff scale is from the $h$-$h$-$h$ vertex, where $h$ is the lighter \textit{CP}-even Higgs boson at a mass below 125 GeV.   The gluon fusion processes of $gg \to h h $ and $gg \to AA$ are insensitive to the cutoff scale, yielding a small variation of the production cross sections, $\mathcal{O}(1)\,{\rm fb}$, according to $\Lambda_{\rm cut}$. The smoking-gun signature is from the triple Higgs production of $q\bar{q}' \to W^* \to H^\pm hh$, which solely depends on the $h$-$h$-$h$ vertex. The cross section for $\Lambda_{\rm cut}=1\,{\rm TeV}$ is about $10^3$ times larger than that for the Planck-scale cutoff. Since the decay modes of $H^\pm \to W^* h/W^* A$ and $h/A \to bb$ are dominant, the process yields the $6b+\ell\nu$ final state, which enjoys an almost background-free environment. Consequently, the precision measurement of $pp \to H^\pm hh$ can probe the cutoff scale of the model.
\end{abstract}


\maketitle
\tableofcontents

\section{Introduction}
Up to today, all of the measurements of the production cross sections
of the standard model (SM) particles at high-energy colliders are in good agreement with the SM predictions~\cite{Ellis:2021kzk},
including  the observed Higgs boson with a mass of $125\gev$~\cite{ATLAS:2020fcp,ATLAS:2020bhl,CMS:2020zge,ATLAS:2021nsx,CMS:2021gxc,ATLAS:2020syy,ATLAS:2021upe,ATLAS:2020pvn,CMS:2021ugl,ATLAS:2020wny,ATLAS:2020rej,ATLAS:2020qdt,ATLAS:2020fzp,CMS:2020xwi,ATLAS:2021zwx}.
Nevertheless, our minds are rarely in satisfaction with the SM,
because of the unsolved questions such as the naturalness problem, baryogenesis, non-zero neutrino masses,
fermion mass hierarchy, the origin of \textit{CP} violation in the quark sector,  
and the identity of dark matter.
We continue our journey in the quest for the ultimate theory. 

A crucial question is whether the ultimate theory shall reveal its whole structure
at one energy scale.
The answer is much more likely to be \emph{no}
when looking back on the SM, the only reliable guideline at this moment.
We witnessed the emergence of SM particles in stages.
The same phenomena could happen in the ultimate theory.
In other words, the final theory may consist of multilevel sub-models. 
The first-stage NP model\footnote{Two different structures exist for the first-stage NP model. 
It can take over the SM from a high-energy scale,
accommodating new heavy particles with the multi-TeV masses.
Or it coexists with the SM at the electroweak scale
so that the new particles have the intermediate masses.
}
describes our universe up to a particular energy scale $\lmc$
and then hands over its role to the second-stage NP model.
$\lmc$ could be as high as the Planck scale 
or as low as $10\tev$.
Then can an observable distinguish the high and low $\lmc$?
This is the driving question in our paper.

For the first-stage NP model,
we consider the two-Higgs-doublet model (2HDM)
since many fundamental questions are closely related to the Higgs sector.
The 2HDM provides the answers to some questions.
For example, 
the first-order electroweak phase transition in the 2HDM can explain the baryogenesis~\cite{Turok:1990zg,Cohen:1991iu,Zarikas:1995qb,Cline:1996mga,Fromme:2006cm}.
However, the model cannot address all the fundamental questions,
which makes it a suitable candidate for the first-stage NP model.

Then the next question is how to calculate the energy scale at which the second-stage NP model appears.
A good way is to calculate the cutoff scale of the 2HDM.
Even though the theoretical requirements 
(unitarity, perturbativity, and vacuum stability)  are satisfied at the electroweak scale,
they can be broken at a higher energy scale $\lmc$ because 
the parameters evolve under renormalization group equations (RGE)~\cite{Machacek:1983tz,Machacek:1983fi,Machacek:1984zw,Luo:2002ti,Das:2015mwa,Haber:1993an,Grimus:2004yh}.
Since it implies the advent of the second-stage NP model, we call $\lmc$ the cutoff scale of the model.

\begin{figure}[h!]
\centering
\includegraphics[width=0.65\textwidth]{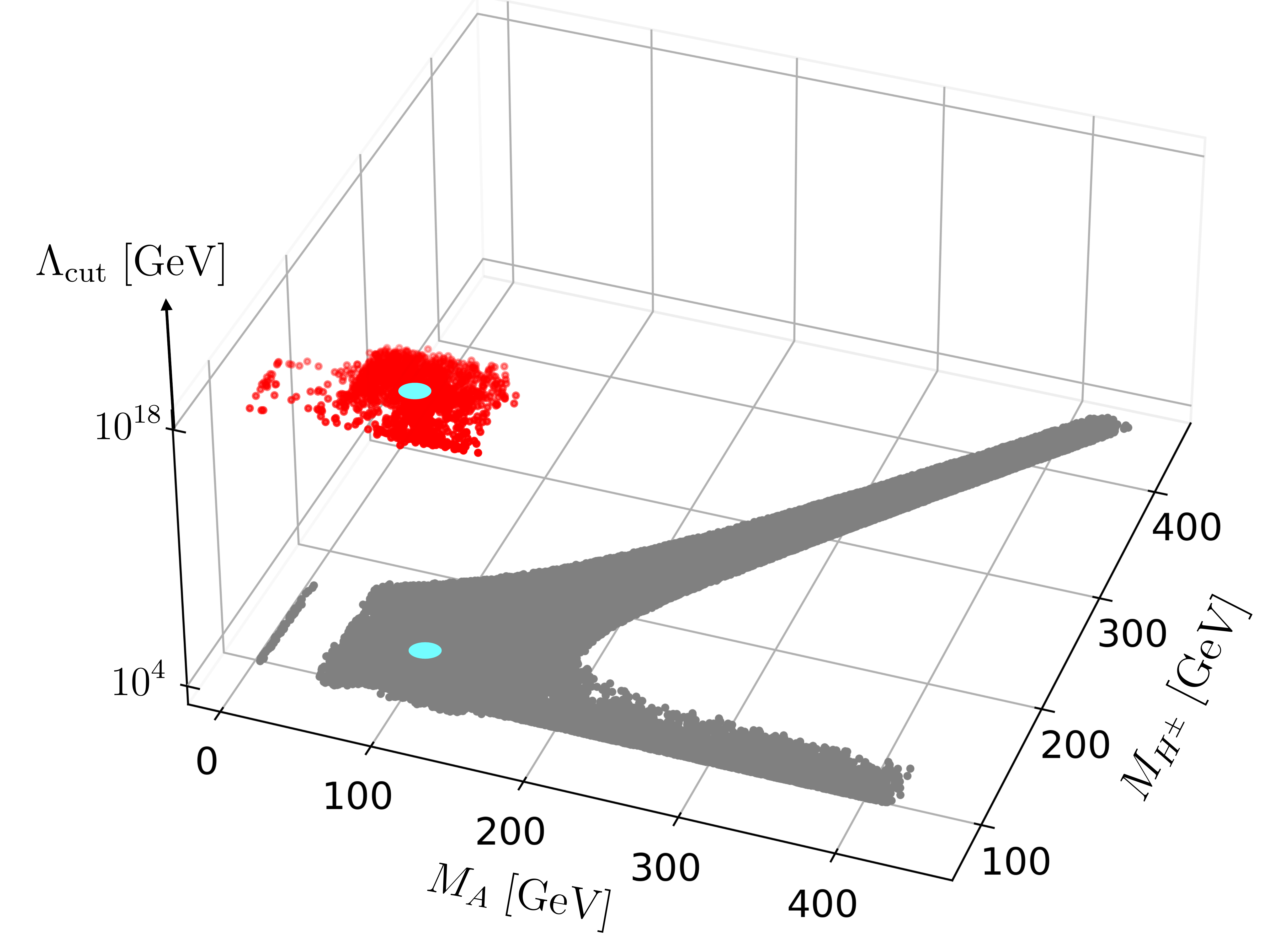}
\caption{
Allowed $(\ma,\mch)$ with the cutoff scales of $\lmc=10\tev$ (gray points) and $\lmc= 10^{18}\gev$ (red points)
in the inverted type I, over the viable parameter points that satisfy the theoretical requirements and the experimental constraints.
Two points in cyan yield the same $\ma$ and $\mch$.
}
\label{fig-intro}
\end{figure}

In the literature,
the high-energy scale behavior of the 2HDM has been extensively studied.
Most studies are focused on the impact of high $\lmc$ on the extra Higgs boson masses~\cite{Das:2015mwa,Chowdhury:2015yja,Basler:2017nzu,Krauss:2018thf,Oredsson:2018yho,Aiko:2020atr,Dey:2021pyn,Lee:2022gyf,Kim:2022hvh,Kim:2022nmm}.
However, observing the scalar mass spectrum that high $\lmc$ predicts 
does not guarantee that $\lmc$ is high.
In this paper, we focus on the type I of the 2HDM in the inverted Higgs scenario 
where the heavier \textit{CP}-even scalar $H$ is the observed Higgs boson~\cite{Ferreira:2012my,Chang:2015goa,Jueid:2021avn}.
The model can accommodate the cutoff scale all the way up to the Planck scale~\cite{Lee:2022gyf}.
As illustrated in \fig{fig-intro},
the allowed points of $(\ma,\mch)$ for $\lmc = 10^{18}\gev$ (red points)
are overlapped with those for $\lmc=10\tev$ (gray points).
Two points in cyan, one for $\lmc=10\tev$ and the other for $\lmc = 10^{18}\gev$, 
have the same $\ma$ and $\mch$.
We need an alternative observable to disentangle the high- and low-cutoff scales.
The measurement of $\tan\beta$,
the ratio of two vacuum expectation values of two Higgs doublet fields, 
is tricky in type I when $\tan\beta$ is large.
In type I, all the Yukawa couplings of the BSM Higgs bosons
are inversely proportional to $\tan\beta$ in the Higgs alignment limit.
So, the main production channels of $gg\to A/h$ and $t\to \ch b$ are suppressed by large $\tan\beta$.
In addition, the branching ratios of all the fermionic decay modes are insensitive to $\tan\beta$ in type I.
We need alternative observables to probe the cutoff scale.

We will show that the trilinear Higgs couplings can play the role.
In particular, the value of the $h$-$h$-$h$ vertex, $\lm_{hhh}$, is highly sensitive to the cutoff scale.
To probe the trilinear Higgs couplings at the LHC,
we will study the di-Higgs processes of $gg\to hh$ and $gg\to AA$,
and the tri-Higgs processes of $pp\to \ch hh$.
The gluon fusion production of $gg\to h/H \to hh/AA$
is to be shown insensitive to $\lmc$ because of the destructive interference between the $h$ and $H$ contributions.
We will show that the tri-Higgs process $pp\to \ch hh$ is the best to measure the cutoff scale
because the signal rate is solely dependent on $\lm_{hhh}$, yielding $\sg_{\lmc=1\tev}/\sg_{\lmc = 10^{18}\gev} \sim 10^3$.
Considering the dominant decays modes of $\ch$ and $h$,
we will suggest for the first time the $6b+\ell\nu$ final state as an efficient discriminator  
between the high- and low-cutoff scales.
These are our new contributions.

The paper is organized in the following way. 
In Sec.~\ref{sec:review}, we briefly review the type I in the inverted scenario.
Section \ref{sec:viable} describes the methods of the parameter scanning and the calculation of $\lmc$.
The characteristics of the viable parameter points with high $\lmc$ are also presented.
In Sec.~\ref{sec:trilinear},
we calculate the correlation between the trilinear Higgs couplings and the cutoff scale.
Section \ref{sec:LHC} deals with the LHC phenomenology.
Based on the study of the branching ratios of the extra Higgs bosons,
we will suggest efficient observables to distinguish the high- and low-cutoff scales.  
Finally we conclude in Sec.~\ref{sec:conclusions}.  

\section{Review of type I of 2HDM in the inverted scenario}
\label{sec:review}
The 2HDM introduces two $SU(2)_L$ complex scalar doublet fields with hypercharge $Y=1$,
$\Phi_1$ and $\Phi_2$~\cite{Branco:2011iw}:
\bea
\label{eq:phi:fields}
\Phi_i = \left( \begin{array}{c} w_i^+ \\[3pt]
\dfrac{v_i +  \rho_i + i \eta_i }{ \sqrt{2}}
\end{array} \right), \quad ( i=1,2)
\eea
where $v_{1}$ and $v_2$ are the nonzero vacuum expectation values of $\Phi_{1}$
and $\Phi_2$, respectively.
The ratio of $v_2/v_1$ defines the mixing angle $\bt$ via $\tan\beta=v_2/v_1$.
For notational simplicity,
we use $s_x=\sin x$, $c_x = \cos x$, and $t_x = \tan x$ in what follows.
The combination of $v_1$ and $v_2$, $v =\sqrt{v_1^2+v_2^2}=246\gev $,
spontaneously breaks the electroweak symmetry.
To prevent flavor-changing neutral currents at tree level,
we impose a discrete $Z_2$ symmetry
under which $\Phi_1 \to \Phi_1$ and $\Phi_2 \to -\Phi_2$~\cite{Glashow:1976nt,Paschos:1976ay}.
We allow softly broken $Z_2$ symmetry
since it does not affect the RGE of the dimensionless quartic couplings~\cite{Haber:2006ue}:
the hard $Z_2$ breaking in the Yukawa sector causes too fast growth
of the scalar quartic couplings in the RG running~\cite{Oredsson:2018yho}.

For simplicity, we employ a \textit{CP}-conserving scalar potential that softly breaks the $Z_2$ symmetry, which is given by
\bea
\label{eq:VH}
V_\Phi = && m^2 _{11} \Phi^\dagger_1 \Phi_1 + m^2 _{22} \Phi^\dagger _2 \Phi_2
-m^2 _{12} ( \Phi^\dagger_1 \Phi_2 + \hc) \\ \nn
&& + \frac{1}{2}\lambda_1 (\Phi^\dagger _1 \Phi_1)^2
+ \frac{1}{2}\lambda_2 (\Phi^\dagger _2 \Phi_2 )^2
+ \lambda_3 (\Phi^\dagger _1 \Phi_1) (\Phi^\dagger _2 \Phi_2)
+ \lambda_4 (\Phi^\dagger_1 \Phi_2 ) (\Phi^\dagger _2 \Phi_1) \\ \nn
&& + \frac{1}{2} \lambda_5
\left[
(\Phi^\dagger _1 \Phi_2 )^2 +  \hc
\right],
\eea
where the $m_{12}^2$ term softly breaks the $Z_2$ symmetry.
The scalar potential $V_\Phi$ yields five physical Higgs bosons, the lighter \textit{CP}-even scalar $h$,
the heavier \textit{CP}-even scalar $H$, the \textit{CP}-odd pseudoscalar $A$,
and a pair of charged Higgs bosons $H^\pm$.
Relations of mass eigenstates with weak eigenstates in terms of two mixing angles of $\al$ and $\bt$ are referred to Ref.~\cite{Song:2019aav}.
The SM Higgs boson is a linear combination of $h$ and $H$, given by
\bea
\label{eq:hsm}
\hsm = \sba \,h + \cba \,H.
\eea

The observed Higgs boson at a mass of $125\gev$ at the LHC~\cite{ATLAS:2020fcp,ATLAS:2020bhl,CMS:2020zge,ATLAS:2021nsx,CMS:2021gxc,ATLAS:2020syy,ATLAS:2021upe,ATLAS:2020pvn,CMS:2021ugl,ATLAS:2020wny,ATLAS:2020rej,ATLAS:2020qdt,ATLAS:2020fzp,CMS:2020xwi,ATLAS:2021zwx} has so far agreed with the predictions for the SM Higgs boson.
The SM-like Higgs boson strongly motivates the Higgs alignment limit in the 2HDM.
Two scenarios exist for the limit,
the normal scenario where $\hsm=h$ (i.e., $\sba=1$)
and the inverted scenario where $\hsm=H$ (i.e., $\cba=1$).
In this paper, we concentrate on the inverted scenario in the Higgs alignment limit:
\bea
\label{eq:model:specify}
\mhh=125\gev,\quad \cba=1.
\eea
Then we have the following five parameters:
\bea
\label{eq:model:parameters}
\{ \mh,\quad \ma, \quad \mch, \quad \tb, \quad m_{12}^2 \},
\eea
which define one parameter point.
Then, the quartic coupling constants in $V_\Phi$ are written as~\cite{Kanemura:2011sj}
\bea
\label{eq:lm1}
\lm_1 &=& \frac{1}{v^2}
\left[
m_{125}^2 + \tb^2 \lf \mh^2 - M^2 \ri
\right],
\\ \nn
\lm_2 &=& \frac{1}{v^2}
\left[
m_{125}^2 + \frac{1}{\tb^2} \lf \mh^2 - M^2 \ri
\right],
\\ \nn
\lm_3 &=& \frac{1}{v^2}
\left[
m_{125}^2 - \mh^2 -M^2 +2 \mch^2 
\right],
\\ \nn
\lm_4 &=& \frac{1}{v^2}
\left[
M^2+\ma^2-2 \mch^2
\right],
\\ \nn
\lm_5 &=& \frac{1}{v^2}
\left[
M^2-\ma^2
\right],
\eea
where $m_{125}=125\gev$ and $M^2 = m_{12}^2/(\sb \cb)$.

The Yukawa couplings to the SM fermions are parametrized as 
\bea
\label{eq:Lg:Yukawa}
\lg^{\rm Yuk} &=&
- \sum_f 
\lf 
\frac{m_f}{v} \xi^h_f \bar{f} f h + \frac{m_f}{v} \kp_f^H \bar{f} f H
-i \frac{m_f}{v} \xi_f^A \bar{f} \gm_5 f A
\ri
\\ \nn &&
- 
\left\{
\dfrac{\sqrt{2}}{v } \overline{t}
\left(m_t \xi^A_t \text{P}_- +  m_b \xi^A_b \text{P}_+ \right)b  H^+
+\dfrac{\sqt m_\tau}{v}\xi^A_\tau \,\overline{\nu}_\tau P_+ \tau H^+
+\hc
\right\},
\eea
which are different according to the 2HDM type.
In this work, we focus on type I.
To facilitate the discussion below,
we will call the type I with the conditions of \eq{eq:model:specify} the inverted type I.
Then the Higgs coupling modifiers are
\bea
\label{eq:xi}
\xi^H_f =1, \quad \xi^h_{t,b,\tau}=\frac{1}{\tb}, \quad \xi^A_{t} = -\xi^A_{b,\tau} = \frac{1}{\tb}.
\eea

The trilinear Higgs couplings as dimensionless parameters are defined by 
\bea
\lg_{\rm tri} &=&
\sum_{\varphi_0=h,H} v \left\{
\frac{1}{3!} \lmh_{\varphi_0\varphi_0\varphi_0 } \varphi_0^3
+\frac{1}{2} \lmh_{\varphi_0 AA  } \, \varphi_0  A^2 
+ \lm_{\varphi_0 H^+ H^-} \,\varphi_0 H^+ H^- \right\}
 \\[3pt] \nn && ~
+ \frac{1}{2} \lmh_{H hh}\, v H h^2  + \frac{1}{2} \lmh_{hHH} \, v h H^2
.
\eea
In the inverted type I, the couplings are~\cite{Bernon:2015qea,Bernon:2015wef}
\begin{align}
\label{eq:trilinear:couplings}
\lmh_{HHH} &= -\frac{3\mhsm^2}{v^2}, \qquad \lmh_{ hHH} = 0,
\\[5pt] \nn
\lmh_{hhh} &=3 \lmh_{hAA} = 3\lmh_{h H^+ H^-} = -\frac{3 (M^2-\mh^2)(\tb^2-1)}{\tb v^2},
\\[5pt] \nn
\lmh_{Hhh} &=  -\frac{\mhsm^2+2 \mh^2-2 M^2}{v^2},
\\[5pt] \nn
\lmh_{HAA} &=
-\frac{\mhsm^2+2\ma^2-2 M^2}{v^2},
\\[5pt] \nn
\lmh_{H H^+ H^-} &=
-\frac{\mhsm^2+2\mch^2-2M^2}{v^2}.
\end{align}
The Higgs alignment limit makes the trilinear coupling of the observed Higgs boson $H$ the same as in the SM,
$\lmh_{HHH} \simeq 0.77$, which is one of the most important targets to measure
at the HL-LHC and future colliders~\cite{Plehn:1996wb,Djouadi:1999rca,Barger:2013jfa,Dawson:2015oha,Cheung:2020xij,Jueid:2021qfs}.
Another remarkable feature is that 
$\lmh_{hhh}$, $\lmh_{hAA}$, and $\lmh_{h H^+ H^-}$
have the common factor of $\tb(M^2-\mh^2)$ in the large $\tb$ limit.

\section{Scanning and RGE analysis}
\label{sec:viable}

Before studying the high-energy behavior of the model via RGE,
the preparation of the allowed parameter points at the electroweak scale is an essential prerequisite.
Therefore,
we randomly scan five model parameters in the range of
\bea
\label{eq:scan:range}
\tb &\in& [1,50], \quad \ma \in [10,3000]\gev, \quad m_{12}^2 \in [-3000^2,3000^2]\gev^2, 
\\ \nn
\mch &\in& [80,3000]\gev, \quad \mh   \in [10,120]\gev ,
\eea
and cumulatively impose the following constraints:
\renewcommand\labelenumi{(\theenumi)}
\bit
\item \textbf{Theoretical requirements}\\
We demand the bounded-from-below Higgs potential~\cite{Ivanov:2006yq},
tree-level unitarity of scalar-scalar scatterings~\cite{Ginzburg:2005dt,Branco:2011iw,Kanemura:2015ska,Arhrib:2000is},
perturbativity~\cite{Chang:2015goa}, and the stability of the \textit{CP}-conserving vacuum
 with $v=246\gev$~\cite{Ivanov:2008cxa,Barroso:2012mj,Barroso:2013awa}.
We use the public code \textsc{2HDMC}-v1.8.0~\cite{Eriksson:2009ws}.
For the perturbativity, \textsc{2HDMC}
requires that the magnitudes of all the quartic couplings among physical Higgs bosons be less than $4\pi$.
However, \textsc{2HDMC} does not check whether our vacuum is the global minimum of the potential.
We demand the tree-level vacuum stability condition of~\cite{Barroso:2013awa}
\bea
m_{12}^2 \lf m_{11}^2 - k^2 m_{22}^2 \ri \lf \tb-k\ri >0 ,
\eea
where $k=(\lm_1/\lm_2)^{1/4}$.
The tree-level conditions have been known to be more than sufficient 
up to very high scales in the Higgs alignment limit~\cite{Basler:2017nzu,Branchina:2018qlf}.
\item  \textbf{Peskin-Takeuchi oblique parameters}~\cite{Peskin:1991sw}\\
The oblique parameters of $S$, $T$, and $U$ in the 2HDM~\cite{He:2001tp,Grimus:2008nb,Zyla:2020zbs}
should satisfy the current best-fit results at 95\% C.L.~\cite{ParticleDataGroup:2022pth}:
	\begin{align}
	\label{eq:STU:PDG}
	S &= -0.02 \pm 0.10,
	\\ \nn
	T &= 0.03 \pm 0.12, \quad 
	U=0.01 \pm 0.11, 
	\\ \nn
	 \rho_{ST} &= 0.92, \quad \rho_{SU}=-0.80,\quad \rho_{TU}=-0.93,
	\end{align}
	where $\rho_{ij}$ is the correlation matrix.
\item  \textbf{Flavor changing neutral currents}\\
We demand that the most recent observables from $B$ physics 
	 be satisfied at 95\% C.L.~\cite{Arbey:2017gmh,Sanyal:2019xcp,Misiak:2017bgg}. 
	 We adopt the results of Ref.~\cite{Arbey:2017gmh}.
\item  \textbf{Higgs precision data} \\ 
We use the public code \textsc{HiggsSignals}-v2.6.2~\cite{Bechtle:2020uwn} to check
the consistency with the Higgs precision data.
Based on the $\chi^2$ value for 111 Higgs
observables~\cite{Aaboud:2018gay,Aaboud:2018jqu,Aaboud:2018pen,Aad:2020mkp,Sirunyan:2018mvw,Sirunyan:2018hbu,CMS:2019chr,CMS:2019kqw} with five parameters,
we require that the $p$-value be larger than 0.05.
\item  \textbf{Direct search bounds}\\
We demand that the model prediction to the cross sections of the direct search modes for new scalar bosons  
at the LEP, Tevatron, and LHC should be less than 95\% C.L. upper bound on the observed cross sections.
The open code \textsc{HiggsBounds}-v5.10.2~\cite{Bechtle:2020pkv} is used.
\eit

Brief comments on the recent CDF measurement 
of the $W$-boson mass~\cite{CDF:2022hxs}, $m_W^{\rm CDF} = 80.4335 \pm 0.0094\gev$,
are in order here.
If we accept $m_W^{\rm CDF}$,
the oblique parameters change into $S_{\rm CDF}=0.15\pm 0.08$ and $T_{\rm CDF}=0.27\pm 0.06$ with $U=0$~\cite{Lu:2022bgw}.
Although $m_W^{\rm CDF}$ has important implications on the 2HDM~\cite{Fan:2022dck,Zhu:2022tpr,Lu:2022bgw,Zhu:2022scj,Song:2022xts,Bahl:2022xzi,Heo:2022dey,Babu:2022pdn,Biekotter:2022abc,Ahn:2022xeq,Han:2022juu,Arcadi:2022dmt,Ghorbani:2022vtv,Broggio:2014mna,Lee:2022gyf,Kim:2022hvh},
our main conclusion on the role of trilinear Higgs couplings in disentangling the high- and low-cutoff scales
does not change. 
Therefore, we focus on the oblique parameters without the CDF $m_W$ measurement. 

Over the parameter points that pass the above constraints,
we evolve the following parameters via the RGE, 
by using the public code \textsc{2HDME}~\cite{Oredsson:2018yho,Oredsson:2018vio}:
\bea
\label{eq:running:parameters}
g_{1,2,3}, \quad \lm_{1,\cdots,5}, \quad \xi^{h,H,A}_f,\quad m_{11}, \quad m_{12},
\quad m_{22}^2, \quad v_{1,2}.
\eea
We include the mixing effects of two scalar doublet fields (with equal quantum numbers) on $v_1$ and $v_2$, which bring about the RG running of $\tb$.
The top quark pole mass scale, $m_{t}^{\rm pole} = 173.4\gev$,
is used to match the 2HDM to the SM.
The boundary conditions at $m_{t}^{\rm pole}$ are referred to Ref.~\cite{Oredsson:2018yho}.
The $\bt$ functions of the gauge, Yukawa, and quartic couplings at one loop level
are presented in Appendix \ref{appendix:RGE}.
Since the difference between the one-loop and two-loop RG running is not large,\footnote{For randomly selected parameter points,
we compared $\lmc$ at one-loop level with $\lmc$ at two-loop level.
The difference is $\mathcal{O}(10)\%$.}
we take the one-loop RGE to efficiently cover all the parameter points.

Now let us describe how we obtained the cutoff scale $\lmc$.
For each parameter point,
we perform the RGE evolution up to the next high energy scale\footnote{To cover from the electroweak scale
to the Planck scale, we take a uniform step in $\log(Q)$.}
and check three conditions, tree-level unitarity, perturbativity, and vacuum stability.\footnote{Note that \textsc{2HDME} use the perturbativity condition of $|\lm_{1,\cdots,5}|<4\pi$ and the tree-level vacuum stability conditions in Refs.~\cite{Ivanov:2006yq,Ivanov:2007de}.}
If all three are satisfied,
we increase the energy scale into the next step. 
If any condition is violated,
we stop the running and record the energy scale as $\lmc$.
We find that the Landau pole appears at a higher scale than $\lmc$.
We additionally require that the cutoff scale should be larger than $1\tev$.
In what follows,
the \enquote{viable parameter points} denote the parameter points that satisfy the aforementioned constraints
and $\lmc>1\tev$.

\begin{figure}[t!]
\centering
\includegraphics[width=0.95\textwidth]{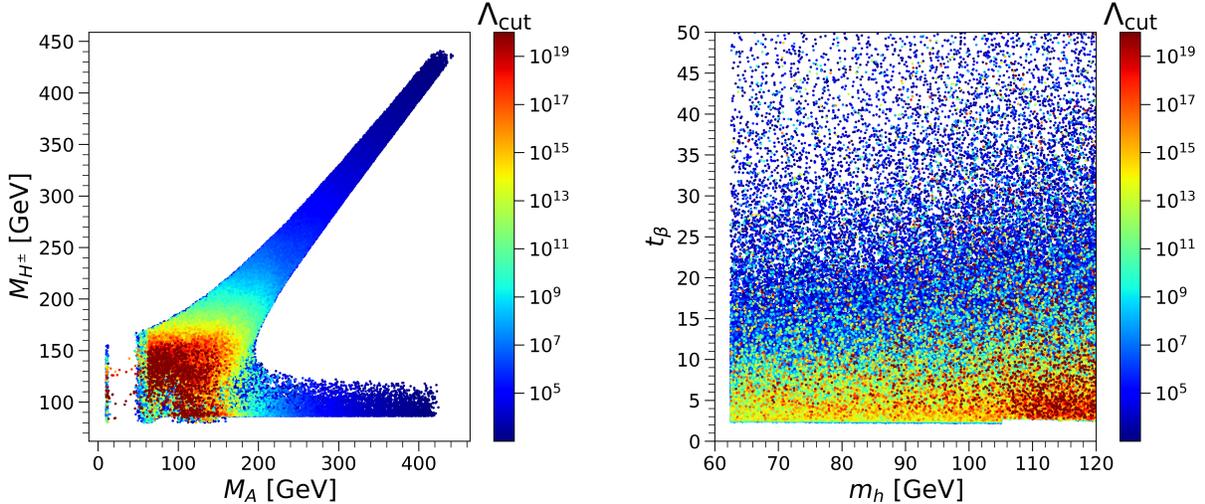}
\caption{
$\mch$ versus $\ma$ in the left panel and $\tb$ versus $\mh$ in the right panel.
The color code denotes the cutoff scale $\lmc$.
}
\label{fig-viable-param-cutoff}
\end{figure}

Strong correlations exist between the cutoff scale and the model parameters.
In Fig.~\ref{fig-viable-param-cutoff},
we present the viable parameter points over the plane of $(\ma,\mch)$ in the left panel
and $(\mh,\tb)$ in the right panel.
The color code denotes the cutoff scale $\lmc$.
We sorted the parameter points according to $\lmc$,
and stacked them in order of $\lmc$,
the points with low $\lmc$ underneath and those with high $\lmc$ on top.
The overlap is due to the projection of five-dimensional parameter space in \eq{eq:model:parameters}
on a two-dimensional subspace.

Several remarkable features are shown in Fig.~\ref{fig-viable-param-cutoff}.
First, the viable parameter points are pretty limited even with the weak condition of $\lmc>1\tev$.
The upper bounds on $\ma$ and $\mch$ exist as $\ma,\mch\lsim 430\gev$,
to which the condition of $\lmc>1\tev$ plays a critical role. 
The Peskin-Takeuchi oblique parameter $T$ is satisfied if $\mch\sim \ma$ or $\mch\sim\mh$,
which explains two branches in the left panel of Fig.~\ref{fig-viable-param-cutoff}.
The lower branch corresponds to $\mch\sim\mh$, which puts the upper bound on the charged Higgs boson mass.
For the intermediate mass range of $\ma,\mch\lsim 200\gev$,
there is no particular correlation between $\ma$ and $\mch$.
However, meaningful correlations appear outside the box with $\ma,\mch\lsim 200\gev$.
If $\mch \gsim 200\gev$ (belonging to the upper branch), $\mch\simeq\ma$.
If $\mch \lsim 200\gev$ and $\ma\gsim 200\gev$ (belonging to the lower branch), $\mch \sim 100\gev$.
The lighter \textit{CP}-even Higgs boson mass $\mh$ (right panel of \fig{fig-viable-param-cutoff})
is heavier than half the observed Higgs boson mass due to the strong constraint from the exotic Higgs boson decay of $H\to hh$.
For $\tb$, most values in the scanning are permitted.
Although the density of the allowed $\tb$ in the scatter plot
is lower for larger $\tb$,
we cannot conclude that large $\tb$ is disfavored: nature chooses just one parameter point.

\begin{figure}[t!]
\centering
\includegraphics[width=0.95\textwidth]{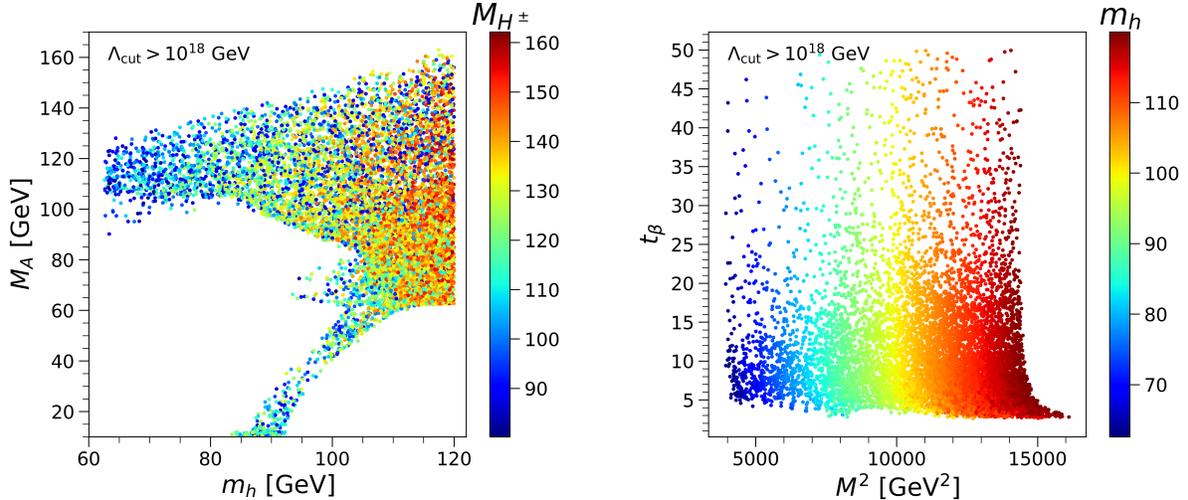}
\caption{
For $\lmc>10^{18}\gev$, $\ma$ versus $\mh$ in the left panel and $\tb$ versus $M^2$ in the right panel.
The color code denotes $\mch$ in the left panel and $\mh$ in the right panel.
}
\label{fig-Planck-viable}
\end{figure}

The second remarkable feature of \fig{fig-viable-param-cutoff} is that substantial parameter points in the inverted type I  remain stable
all the way up to the Planck scale.
Let us investigate the characteristics of the parameter points with the high-cutoff scale.
In \fig{fig-Planck-viable}, we show  $\ma$ versus $\mh$ (left panel) and $\tb$ versus $M^2$ (right panel)
after imposing $\lmc>10^{18}\gev$.
The color code in the left (right) panel denotes $\mch$ ($\mh$).
It is clearly seen that the high-cutoff scale requires \textit{light} masses of the extra Higgs bosons like $\ma, \mch\lsim 160\gev$.
We observe that the charged Higgs boson is lighter than the top quark for the Planck-scale cutoff.
However, the lower bounds of $\ma\gsim 10\gev$, $\mh\gsim 62.5\gev$, and $\mch \gsim 80\gev$ do not change 
by imposing $\lmc>10^{18}\gev$.
Similarly, the Planck-scale cutoff does not affect the values of $\tb$, 
as shown in the right panel of \fig{fig-Planck-viable}.

The values of $M^2$ are also limited even with $\lmc>1\tev$.
Only the positive values of $M^2$ are permitted because negative $M^2$ enhances $\lm_1$ 
through the terms proportional to $\tb^2$.
Large $\lm_1$ at the electroweak scale quickly evolves into an unacceptably large value,
which endangers the global minimum condition of the vacuum~\cite{Ivanov:2008cxa,Barroso:2012mj,Barroso:2013awa}.
For $\lmc>10^{18}\gev$,
a unique correlation of $M^2\simeq \mh^2$ appears as shown in the right panel in \fig{fig-Planck-viable}.
It is also ascribed to the $\tb^2$ terms in $\lm_1$.
The condition of $M^2\simeq \mh^2$ suppresses the $\tb^2$ terms
and thus helps to retain the stability of the scalar potential.


Although the high-cutoff scale demands light masses of the extra Higgs bosons, the inverse is not true.
All the mass spectra in the left panel of \fig{fig-Planck-viable} also accommodate a low-cutoff scale: see \fig{fig-intro}.
Even if we observe $\mh=\ma=100\gev$ and $\mch=140\gev$, for example,
the mass spectrum alone cannot tell whether the cutoff scale is high or low.
The reader may suggest to use $\tb$ as a discriminator of the cutoff scale.
However, measuring $\tb$ in type I is challenging at the LHC, especially when $\tb \gg 1$.
The value of $\tb$ governs the fermionic productions (from the top quark decay or gluon fusion via quark loops)
and fermionic decays of the extra Higgs bosons.
If $\tb$ is large, the fermionic production of the extra Higgs bosons is highly suppressed
because all the Yukawa couplings of the extra Higgs bosons
are inversely proportional to $\tb$ in type I.
The bosonic productions such as $\qq\to Z^* \to Ah$ and $\qq\to W^*\to\ch h/A$~\cite{Akeroyd:1995hg,Akeroyd:1998ui,Barroso:1999bf,Brucher:1999tx,Akeroyd:2003xi,Arhrib:2008pw,Berger:2012sy,Ilisie:2014hea,Delgado:2016arn,Mondal:2021bxa,Cheung:2022ndq,Kim:2022nmm}
do not give information about $\tb$.
Moreover, the fermionic decay parts are insensitive to $\tb$
because of the same dependence of all the Yukawa couplings on $\tb$. 
If we cannot measure the exact value of large $\tb$,
it is reasonable to include all the viable parameter points with $\tb>10$
when pursuing a way to discriminate the high and low $\lmc$.

\section{Trilinear Higgs couplings} 
\label{sec:trilinear}

In this section,
we study the trilinear Higgs couplings to measure $\lmc$.
We consider the following three benchmark points:
\begin{align}
\label{eq:BP}
\hbox{BP-1: }&\quad \mh=70\gev, & \ma&=110\gev, & \mch&=110\gev,
\\ \nn
\hbox{BP-2: }&\quad \mh=100\gev, & \ma&=100\gev,& \mch&=140\gev,
\\ \nn
\hbox{BP-3: }&\quad \mh=110\gev, & \ma&=70\gev,& \mch&=140\gev,
\end{align}
all of which accommodate the cutoff scale from $1\tev$ to $10^{19}\gev$.
As discussed in the previous section,
we focus on the large $\tb$ limit as
\begin{align}
\label{eq:tb:cases}
\hbox{Large $\tb$ case: } & \tb>10.
\end{align} 
For $m_{12}^2$,
we incorporate all the values that satisfy the constraints in Sec.~\ref{sec:viable}.

\begin{figure}[t!]
\centering
\includegraphics[width=\textwidth]{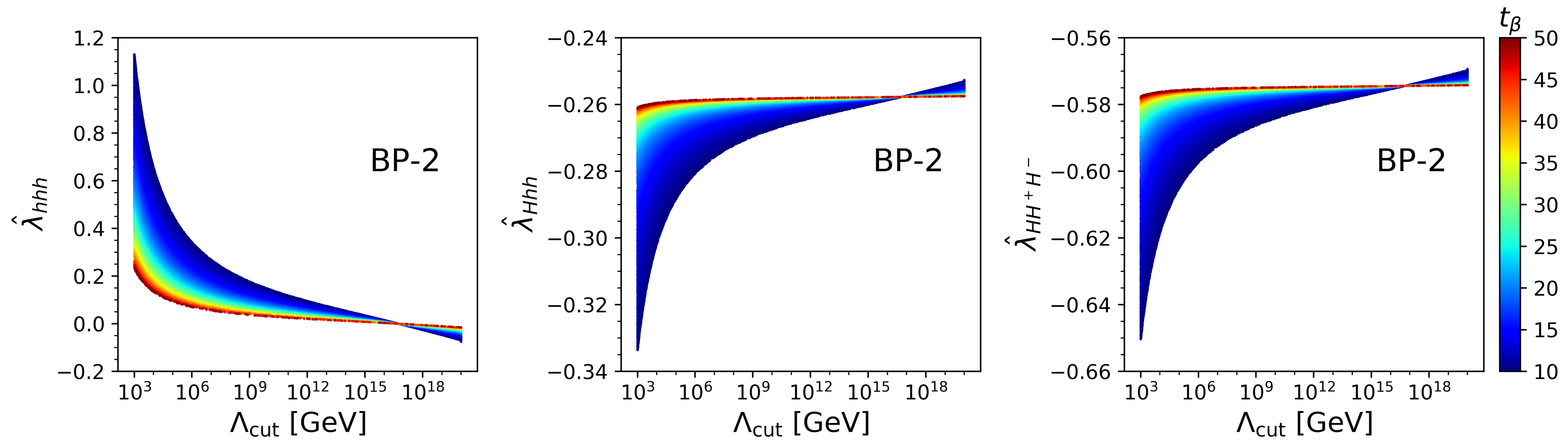}
\caption{
Trilinear Higgs couplings of $\lmh_{hhh}$ (left panel), $\lmh_{Hhh}$ (middle panel),
and $\lmh_{HH^+ H^-}$ (right panel) against the cutoff scale $\lmc$.
The color code denotes $\tb$.
We take BP-2 where $\mh=\ma=100\gev$, $\mch=140\gev$, and $\tb>10$.}
\label{fig-trilinear-cutoff}
\end{figure}

Let us turn into the trilinear Higgs couplings versus $\lmc$.
In \fig{fig-trilinear-cutoff},
we present $\lmh_{hhh}$ (left panel), $\lmh_{Hhh}$ (middle panel),
and $\lmh_{HH^+ H^-}$ (right panel) at the electroweak scale, as a function of $\lmc$:
note that $\lmh_{hAA} = \lmh_{h H^+ H^-} = \lmh_{hhh}/3$. 
Here only the BP-2 results are shown because BP-1 and BP-3 yield similar results with $\mco(10)\%$ differences.
The value of $\tb$ is shown via the color code.
It is impressive that the values of $\lmh_{hhh}$, $\lmh_{Hhh}$, and $\lmh_{HH^+ H^-}$ for $\lmc=10^{18}$
are not overlapped with those for $\lmc \lsim 10^{17}\gev$, 
although the allowed values for lower $\lmc$ are considerably spread by the unfixed $\tb$ and $m_{12}^2$.
The most sensitive dependence on $\lmc$ is shown in $\lmh_{hhh}$,
which ranges in $[-0.09,\, 1.1]$.
Since $\lmh_{hhh}=0$ is included, the change of $\lmh_{hhh}$ according to $\lmc$ is huge.
On the other hand, 
the variations of $\lmh_{Hhh}$ and $\lmh_{HH^+ H^-}$ are small within $10\%\sim 20\%$.

\begin{figure}[t!]
\centering
\includegraphics[width=0.85\textwidth]{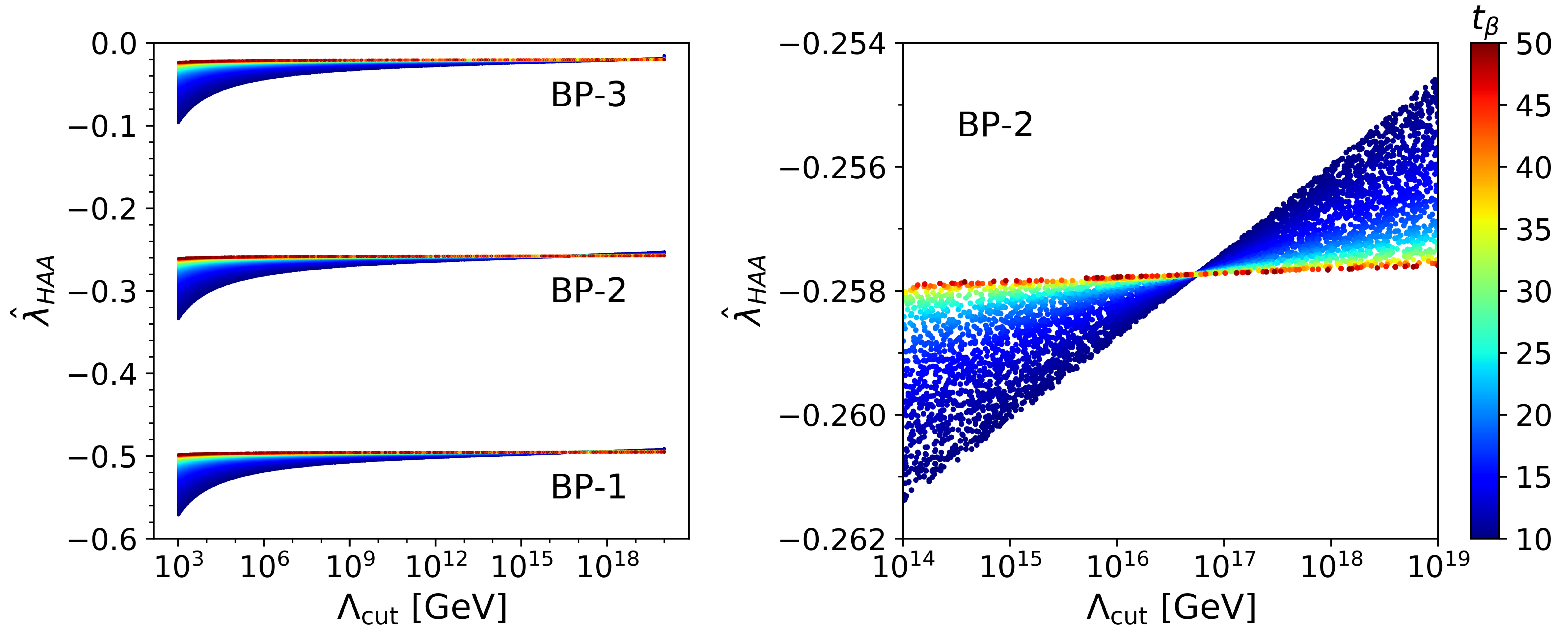}
\caption{
Trilinear Higgs couplings of $\lmh_{HAA}$ versus the cutoff scale $\lmc$
for BP-1, BP-2, and BP-3 (left panel) and $\lmh_{HAA}$ around the \textit{focus} cutoff scale for BP-2 (right panel).
The color code denotes $\tb$.
We include all the viable parameter points with $\tb>10$.}
\label{Fig-tri-HAA}
\end{figure}

%

In the left panel of \fig{Fig-tri-HAA},
we present $\lmh_{HAA}$ versus $\lmc$ for BP-1, BP-2, and BP-3.
Unlike $\lmh_{hhh}$, $\lmh_{Hhh}$, and $\lmh_{HH^+ H^-}$,
the value of $\lmh_{HAA}$ is sensitive to the benchmark point.
For high $\lmc$, $|\lmh_{HAA}|$ of BP-1 is about 25 times larger than that of BP-3.
$\lmh_{HAA}$ depends on $\ma$ and $M^2$, not on $\tb$.
Since $ M^2 \approx \mh^2$ and $\mh \sim \mhsm$ as shown in \fig{fig-Planck-viable},
the $M^2$ contribution to $\lmh_{HAA}$ is nearly cancelled by the $\mhsm$ contribution.
So, the heavier $\ma$ is, the larger $|\lmh_{HAA}|$ is.

\begin{figure}[t!]
\centering
\includegraphics[width=0.8\textwidth]{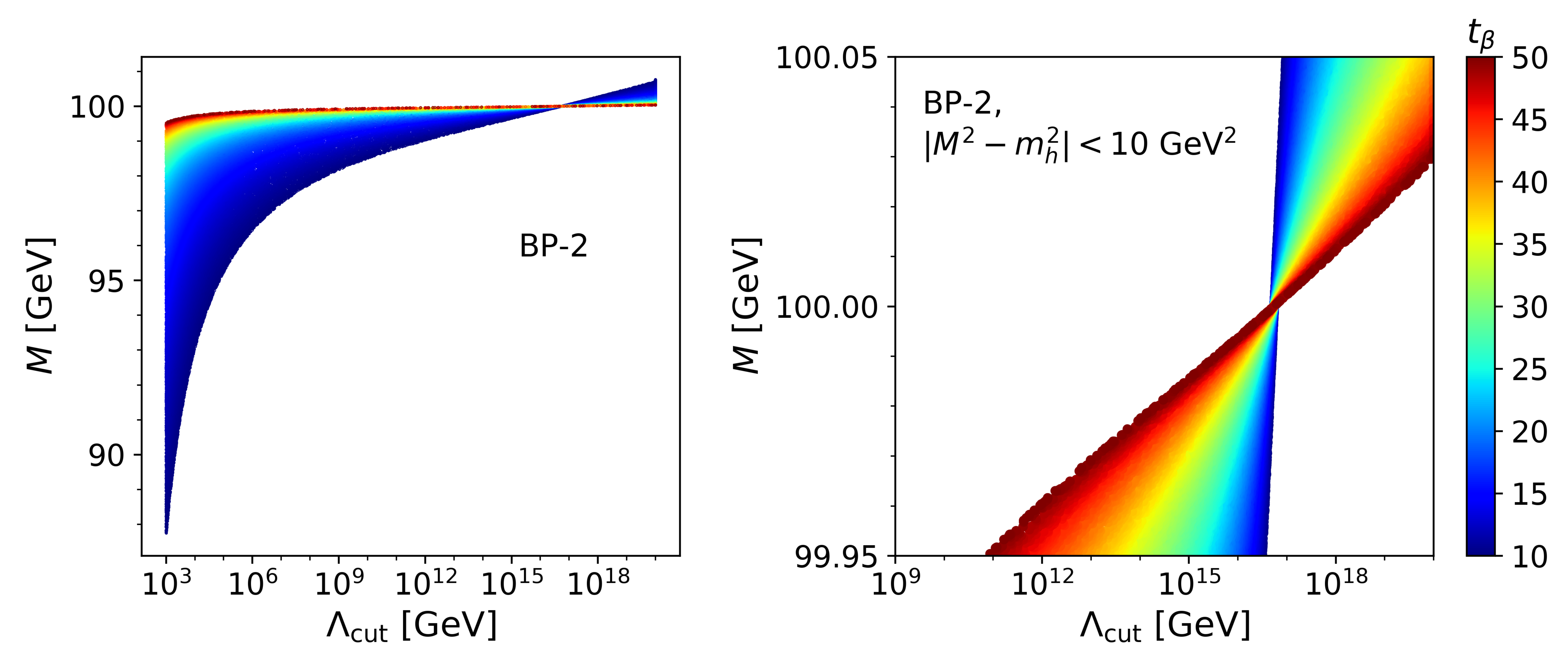}
\caption{
$M(\equiv \sqrt{M^2})$ versus the cutoff scale $\lmc$
for BP-2 with the color code of $\tb$.
The left panel presents the results of all the viable parameter points
and the right panel shows those with $|M^2-\mh^2|\leq 10\gev^2$.
}
\label{Fig-M-cutoff-tb}
\end{figure}

We observe a special $\lmc$ in \fig{fig-trilinear-cutoff} and the right panel of \fig{Fig-tri-HAA}. 
Around $\lmc \simeq 10^{17}\gev$, all the trilinear Higgs couplings are almost fixed.
Even though we present only the results of BP-2 in \fig{fig-trilinear-cutoff},
the same behavior is found in BP-1 and BP-3.
The spread trilinear Higgs couplings 
are focused on a single cutoff point which we call $\lmc^{\rm focus}$.
Since the allowed value of $M^2$ plays a crucial role in understanding $\lmc^{\rm focus}$,
we show $M(\equiv \sqrt{M^2})$ versus the cutoff scale $\lmc$
for BP-2  in \fig{Fig-M-cutoff-tb}.
The left panel presents the results of all the viable parameter points.
It is clear to see that the parameter points with $\lmc=\lmc^{\rm focus}$ satisfy the condition of $M^2 = \mh^2$.
We found that the reverse holds true.
If we allow small deviation like $|M^2-\mh^2|\leq 10\gev^2$ as in the right panel of \fig{Fig-M-cutoff-tb},
the cutoff scale converges to $\lmc=\lmc^{\rm focus}$ for $\tb=10$
but ranges from $10^{12}\gev$ to the Planck scale for $\tb=50$.
Then why does the condition of $M^2 = \mh^2$ fix the trilinear Higgs coupling?
It is because the condition removes the $\tb$ dependence of the trilinear couplings,
which is the main source for their variation.
In addition, $M^2 = \mh^2$ removes the dangerous $\tb^2$ terms of $\lmh_1$,
which guarantees the high-energy scale of $\lmc^{\rm focus}$.

\begin{figure}[t!]
\centering
\includegraphics[width=0.8\textwidth]{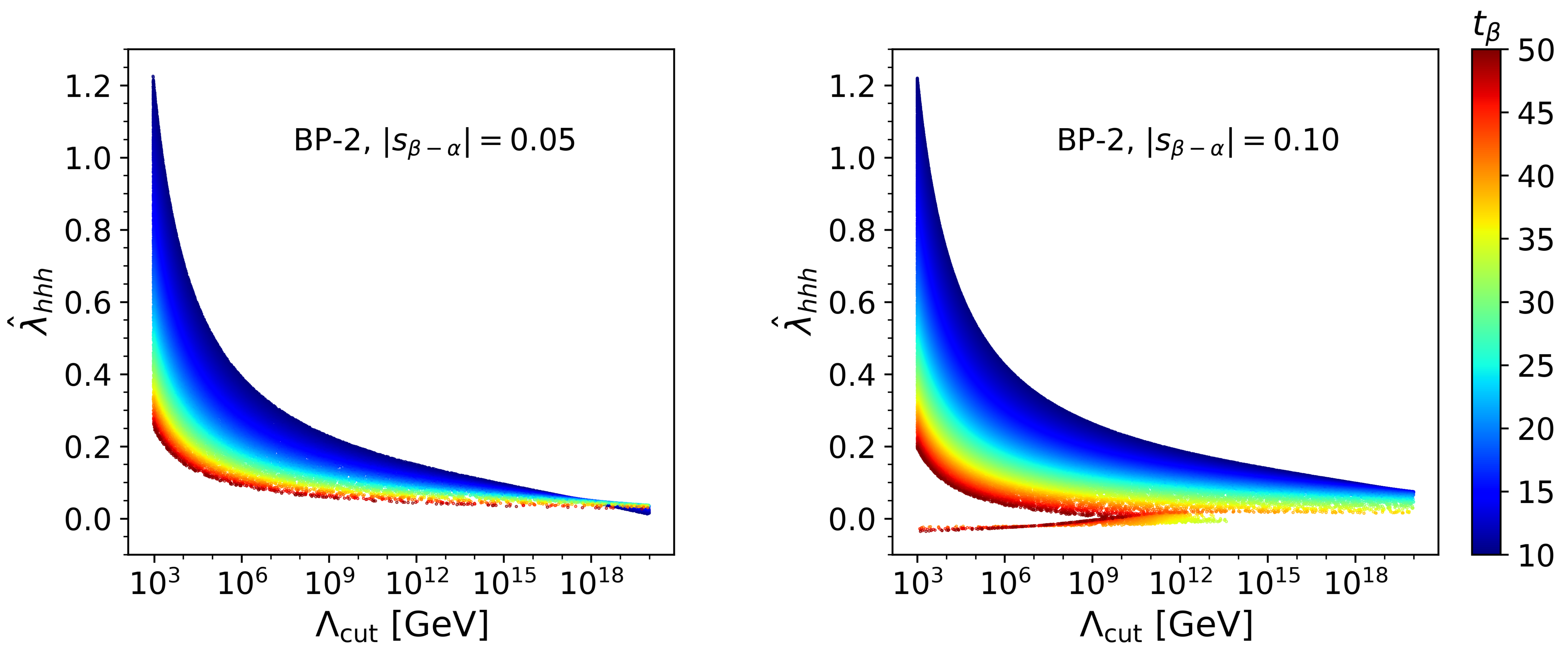}
\caption{
For the small deviation from the Higgs alignment limit, $\lmh_{hhh}$ versus the cutoff scale $\lmc$
for BP-2 with the color code of $\tb$.
The left (right) panel presents the results of $|\sba|=0.05$ ($|\sba|=0.1$).
}
\label{fig-lmhhh-cutoff-NoAlign}
\end{figure}

Now we discuss the other 2HDM scenarios. 
This paper concentrates on the inverted type I in the Higgs alignment limit.
Let us first consider a small deviation from the alignment. 
Since the Higgs precision prospect 
at the future muon collider associated with the HL-LHC and
the Higgs factory at $\sqrt{s}=250\gev$ is $\dt \kp_W \simeq 0.11\%$~\cite{Forslund:2022xjq},
we consider two cases of $|\sba|=0.05,\,0.1$.
In \fig{fig-lmhhh-cutoff-NoAlign}, we present $\lmh_{hhh}$ versus $\lmc$ for $|\sba|=0.05$ (left panel) and 
$|\sba|=0.1$ (right panel).
Here only BP-2 results are shown.
For $|\sba|=0.05$,
the behavior of $\lmh_{hhh}$ about $\lmc$ remains almost same as the alignment case.
The values of $\lmh_{hhh}$ for $\lmc=10^{18}\gev$ are not overlapped with those for $\lmc \lsim 10^{6}\gev$.
If we increase the deviation from the alignment into $|\sba|=0.1$,
the band of $\lmh_{hhh}$ widens.
The values of $\lmh_{hhh}$ for $\lmc=10^{18}\gev$ are mostly overlapped with the low cutoff scale,
except for $\lmc \lsim 10\tev$.

Finally, we discuss whether we have similar results for other types or the normal Higgs scenario where
the lighter \textit{CP}-even Higgs boson is the observed one.
In the inverted Higgs scenario,  type II and type Y are excluded by imposing the condition of $\lmc > 1\tev$~\cite{Lee:2022gyf}
because the constraint from $b\to s\gm$, $\mch>800\gev$~\cite{Misiak:2020vlo},
contradicts the required light masses of the BSM Higgs bosons.
Type X in the inverted scenario can accommodate $\lmc = 10^{18}\gev$
and show similar behaviors of the trilinear Higgs couplings about the cutoff scales.
In the normal scenario,
all four types can retain the theoretical stability up to $\lmc = 10^{18}\gev$.
The high cutoff scale demands the almost exact mass degeneracy among the BSM Higgs boson masses,
i.e., $M=\ma=\mhh=\mch$, but does not put any upper bounds on the masses.
So, the question for the normal scenario is changed:
how can we distinguish the high- and low-cutoff scales via observables 
\textit{if we observe a highly degenerate mass spectrum of the extra Higgs bosons}?
It is more challenging than in the inverted scenario
due to the heavy masses and the soft decay products of the BSM Higgs bosons.
Nevertheless, the similar behavior of $\lmh_{HHH}$ about $\lmc$ 
leaves a motivation for the phenomenological study in future colliders.

\section{LHC phenomenology} 
\label{sec:LHC}

\begin{figure}[t!]
\centering
\includegraphics[width=\textwidth]{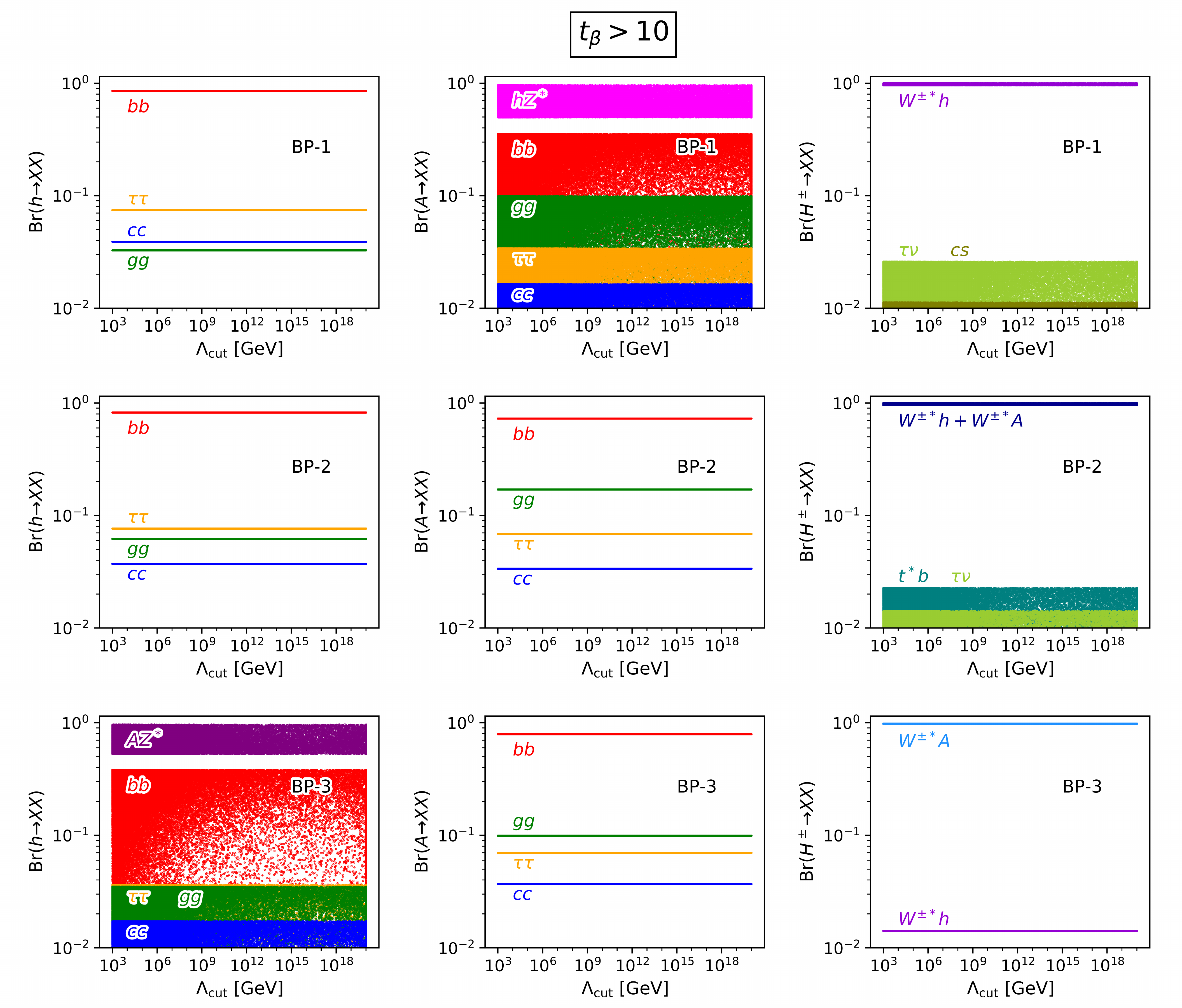}
\caption{
Branching ratios of $h$ (upper panels), $A$ (middle panels), and $\ch$ (lower panels)
 about the cutoff scale $\lmc$.
All the viable parameter points with $\tb>10$ are included.
}
\label{Fig-BR}
\end{figure}

For the LHC phenomenology, we first present the branching ratios of $h$ (left panels), $A$ (middle panels), and $\ch$ (right panels)
as a function of $\lmc$ in \fig{Fig-BR}.
The results of BP-1, BP-2, and BP-3 are in the upper, middle, and lower panels, respectively.
We include all the viable parameter points.
The first noteworthy feature in \fig{Fig-BR} is that 
the dominant decay mode of the extra Higgs boson with the given mass spectra
is insensitive to the cutoff scale,
even though the two parameters of $\tb(>10)$ and $m_{12}^2$ are not fixed.
The decay of $h$ depends on the hierarchy between $\mh$ and $\ma$.
When $\mh\leq \ma$ as in BP-1 and BP-2,
the leading (next-to-leading) decay mode is $h\to bb$ ($h\to \ttau$).
However, if $\mh>\ma$ as in BP-3,
the dominant decay mode is $h \to A Z^*$.
The suppressed Yukawa couplings of $h$ by large $\tb$ enhance the bosonic decays modes
if kinematically open.
The decay of $A$ is primarily determined by the hierarchy between $\ma$ and $\mh$.
For BP-2 and BP-3 with $\ma\leq \mh$, the pseudoscalar $A$ dominantly decays into a pair of $b$ quarks with
$\br(A\to \bb)\gsim 0.73$.
The next-to-leading decay mode of $A$ is into $gg$.
The substantial $\br(A\to gg)$ is attributed to
the larger loop amplitudes of a pseudoscalar than those of a scalar 
for the spin-1/2 particle contributions~\cite{Djouadi:2005gj}.
The third one is $A\to \ttau$.
If $\ma>\mh$ as in BP-1, however, $\br(A \to h Z^*)$ becomes the largest,
which holds for the entire range of $\lmc$.
The decay into $\bb$ is mostly next-to-leading.\footnote{We caution the reader that
the scattered points are overlapped except for the leading decay mode. 
For some parameter points, $A\to gg$ can be next-to-leading.}

The charged Higgs boson mainly decays into $h W^{\pm *}$ and $A W^{\pm *}$,
for the three benchmark points.
Once kinematically allowed, the bosonic decay modes are dominant:
the $\ch$-$\wmp$-$h$ vertex is proportional to $\cba(=1)$;
the $\ch$-$\wmp$-$A$ vertex is originated from the pure gauge interaction.
In BP-3 with $\mch>\ma$ and $\mch>\mh$,
the branching ratios of the fermionic modes are below 1\%.
If either $\mh$ or $\ma$ is beyond the kinematic threshold as in BP-1 and BP-2,
the fermionic decay modes become considerable.
The leading fermionic decay mode depends on the charged Higgs bosons mass.
For BP-1 where $\mch$ is substantially lighter than the top quark mass,
$\ch\to\tau\nu$ has the largest branching ratio among the fermionic decay modes,
followed by $\ch\to c s$.
In BP-2 where $\mch$ is near to the top quark mass,
$\ch\to t^* b$ becomes the leading fermionic mode, followed by $\ch\to\tau\nu$.

\begin{figure}[t!]
\centering
\includegraphics[width=0.8\textwidth]{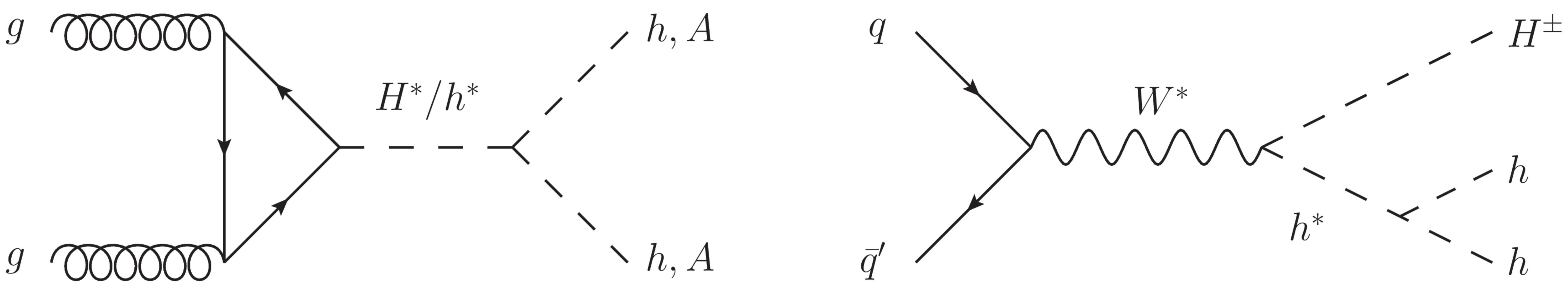}
\caption{
Feynman diagrams of $gg \to hh/AA$ (left panel)
and $q\bar{q}'\to W^* \to \ch hh$ (right panel).}
\label{Fig-Feynman}
\end{figure}

To probe the trilinear Higgs couplings at the LHC,
we need to consider multi-Higgs production mediated by Higgs bosons.
The first important are the di-Higgs processes, $gg\to H/h \to hh/AA$.
The corresponding Feynman diagram\footnote{We omit the box diagrams from the top quark loop
because two factors of $1/\tb$ suppress them.
} is in the left panel of \fig{Fig-Feynman}.
The contribution of $H$ destructively interferes with that of $h$ 
because the sign of $\lmh_{Hhh}$ and $\lmh_{hhh}$ are opposite to each other:
see \fig{fig-trilinear-cutoff}.

\begin{figure}[t!]
\centering
\includegraphics[width=0.9\textwidth]{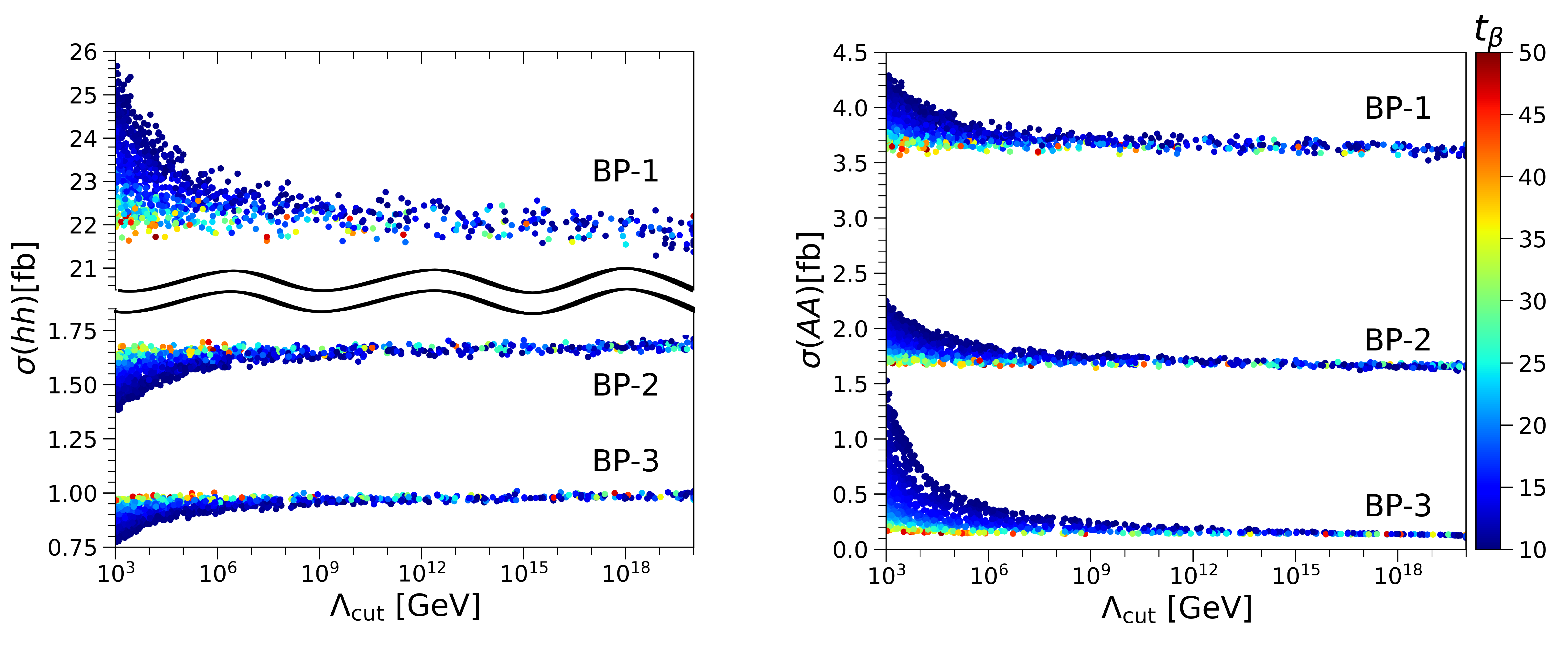}
\caption{
Cross sections of $gg\to hh$ (left panel) and $gg\to AA$  (right panel) at the 14 TeV LHC,
as a function of $\lmc$.
The color code denotes $\tb$.
The description of the benchmarks is in the main text.
}
\label{Fig-DiHiggs-xsection}
\end{figure}

In \fig{Fig-DiHiggs-xsection}, we present as a function of $\lmc$
the parton-level production cross sections
for $gg\to hh$ (left panel) and $gg\to AA$ (right panel) at the 14 TeV LHC over the viable parameter points.
All the three benchmark points in \eq{eq:BP} are considered.
The color code denotes $\tb$.
To calculate the parton-level cross sections,
we first obtained the Universal FeynRules Output (UFO)~\cite{Degrande:2011ua}
by using \textsc{FeynRules}~\cite{Alloul:2013bka}.
Then we interfered the UFO file with \textsc{MadGraph5-aMC@NLO}~\cite{Alwall:2011uj}
and calculated the cross sections at the 14 TeV LHC 
with the \textsc{NNPDF31\_lo\_as\_0118} parton distribution function set~\cite{NNPDF:2017mvq}. 

For the production cross sections of $gg\to hh$, the most crucial factor  is $\mh$.
The lighter $m_h$ is, the larger $\sg(gg \to hh)$ is.
BP-1 yields the largest cross section.
On the contrary,
the production cross section of $gg\to AA$ is larger for heavier $\ma$.
The BP-1, which has the heaviest $\ma$ among the three benchmark points, 
has the largest cross section.
It seems contradictory to the kinematic loss by the heavy $\ma$.
The main reason is that the dominant contribution to $gg\to AA$ is from $H$ and thus
$\lmh_{HAA}$ determines the signal rate.
The heavier $\ma$ is, the larger $\lmh_{HAA}$ is: see \fig{Fig-tri-HAA}.


The dependence of $\sg(g g\to h h/AA)$ on $\lmc$ is not large enough to distinct
the high- and low-cutoff scales.
Even the optimistic case, the process of $g g\to h h$ in BP-1 with $\tb=10$,
makes a few fb difference in the cross sections between $\lmc=1\tev$ and $\lmc=10^{18}\gev$.
It is too small to probe at the HL-LHC with the expected total luminosity of $3\iab$.
For larger $\tb$ like 50,
the di-Higgs production cross sections become more insensitive to $\lmc$.
The weak dependence of $\sg(gg\to hh)$ on $\lmc$
is due to the dominant contribution from $H$ and the destructive interference 
between the $H$ and $h$ contributions.
In the BP-1 with $\tb=10$ and $\lmc=1\tev$, for example,
the cross section from $H$ alone is $\sg(gg\to H\to hh) \simeq 36.5\fb$, 
from $h$ alone is $\sg(gg\to h\to hh) \simeq 1.1\fb$,
and from the interference $\sg(gg\to hh)_{\rm intf} \simeq -12.1\fb$.
The $\lmh_{Hhh}$ controls the cross section, but its variation about $\lmc$ is small.

To single out only one trilinear Higgs coupling,
we consider triple Higgs productions at the LHC.
Since the gluon fusion production of the tri-Higgs process through the top quark loop is suppressed by large $\tb$,
we concentrate on the tri-Higgs productions mediated by the gauge bosons.
Through the $Z$ boson, we have
\bea
\label{eq:LHC:Ahh}
&& \qq \to Z^* \to A h^* \to Ahh,\quad \qq \to Z^* \to A^* h\to Ahh,
\\ \label{eq:LHC:hhh}
&& \qq \to Z^* \to A h^* \to AAA,
\eea
and through $W$
\bea
\label{eq:LHC:cHhh} 
&& q\bar{q}' \to W^* \to \ch h^* \to \ch hh,
\\  \label{eq:LHC:cHAA}
&& q\bar{q}' \to W^* \to \ch h^* \to \ch AA,
\\  \label{eq:LHC:cHAh}
&& q\bar{q}' \to W^* \to \ch A^* \to \ch Ah.
\eea
As a representative,
we present the Feynman diagram of $q\bar{q}' \to \ch hh$ 
in the right panel of \fig{Fig-Feynman}.
Since all the above processes in Eqs.~(\ref{eq:LHC:Ahh})--(\ref{eq:LHC:cHAh}) have the same topology of the Feynman diagram,
the production cross sections as a function of $\lmc$ show almost the same behavior.
In BP-2, for example, $\sg(pp\to Ahh)/\sg(pp\to \ch hh)\simeq 0.9$ holds for all $\lmc$.

\begin{figure}[t!]
\centering
\includegraphics[width=\textwidth]{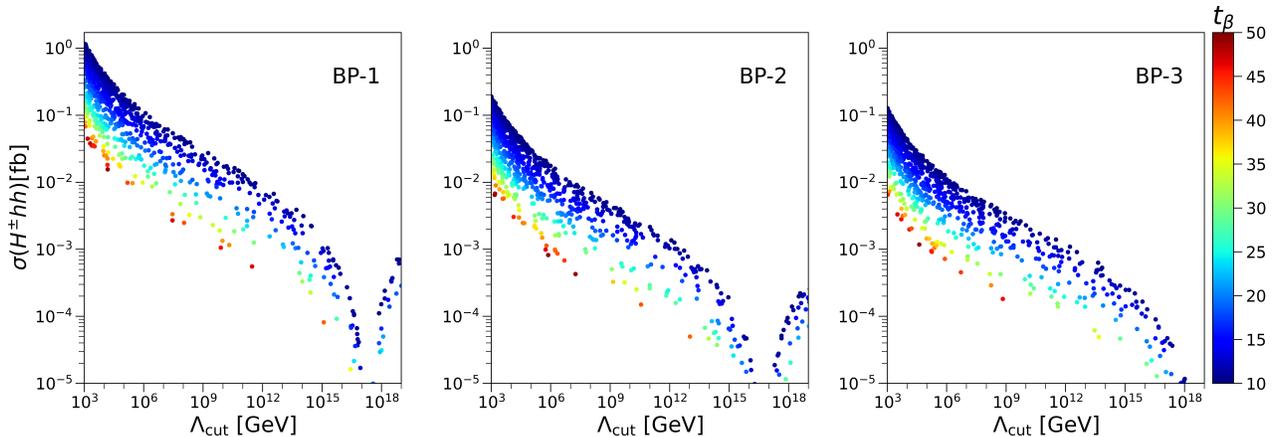}
\caption{
Cross sections of $q\bar{q}' \to \ch hh$ at the 14 TeV LHC
as a function of $\lmc$, for BP-1 (left), BP-2 (middle), and BP-3 (right). 
The color codes indicate $\tb$.
}
\label{Fig-xsection-cHhh}
\end{figure}

In \fig{Fig-xsection-cHhh}, we present 
the parton-level production cross sections
of $q\bar{q}' \to \ch hh$ at the 14 TeV LHC for BP-1 (left panel), BP-2 (middle panel), and BP-3 (right panel).
The color code denotes $\tb$.
The difference of the cross section according to $\lmc$ is big enough to distinguish the high- and low-cutoff scales
of the inverted type I.
The ratio of the cross section for $\lmc=1\tev$ to that for $\lmc =10^{18}\gev$
is more than about $10^3$.
This is the most remarkable result of our study.
Measuring the signal rate of the triple Higgs production tells whether the cutoff scale is high or low.

Finally, we discuss the discovery potential at the HL-LHC.
Discriminating the high- and low-cutoff scales through $gg\to hh/AA$
requires the precision measurement on the cross section within $\sim 1\fb$.
Let us roughly estimate the feasibility.
The processes of $gg\to hh$ and $gg\to AA$ mainly yield $4b$ final states
because $A/h\to bb$ is leading or next-to-leading.
Resembling the di-Higgs process in the SM,
they are challenging to observe for two reasons.
First, the cross section itself is too small.
The maximum cross section, which happens for $\lmc=1\tev$, reaches $\mco(10)\fb$.
It is to be compared with the SM leading order result of $\sg(gg\to \hsm\,\hsm)_{\rm LO} \simeq 17\fb$ at $\sqrt{s}=14\tev$~\cite{Baglio:2012np}.
Since the projected signal significance of the SM Higgs boson pair production at the HL-LHC 
with the total luminosity of $3\ab^{-1}$ is $3.0\,\sigma$,
when the $bbbb$, $bb\ttau$, and $bb\rr$ decay channels are all combined~\cite{ATLAS:2018rvj},
it is hard to observe $gg\to hh/AA$. 
The second difficulty comes from the softer $b$ jets than in the SM di-Higgs process.
For $gg\to hh\to 4b$, 
the lighter $\mh$ than $125\gev$ yields soft $b$ quarks.
For $gg\to AA\to 4b$, 
most of the viable parameter points in \fig{fig-Planck-viable} have $\ma <125\gev$,
which generate soft $b$ quarks. 
As the $b$ tagging efficiency is reduced, the signal significance decreases as well.
In summary, the di-Higgs process is not efficient to probe the cutoff scale.

The triple Higgs production of $\ch hh$ has higher discovery potential.
Since the charged Higgs boson mainly decays into $W^{\pm *}A/h$ in the three benchmark points,
followed by $h/A \to bb$,
the final state is $6b +\ell\nu$. 
This attractive channel has not been studied in the literature.\footnote{The final state of $4j+\ell\nu$ was studied for the vector boson scattering of $WW\to WW$~\cite{Ballestrero:2008gf},
and the SM Higgs boson decaying into a fat jet consisting of $4b/6b/8b$ was studied~\cite{Jung:2021tym}.}
The main backgrounds are 
\begin{align}
t +\bar{t} + \ell\nu &\to b j_{b}^{\rm mis}  j_{b}^{\rm mis} + b  j_{b}^{\rm mis}  j_{b}^{\rm mis} +\ell\nu, 
\\ \nn
t +\bar{t} + jj &\to b \ell\nu + b  j_{b}^{\rm mis}  j_{b}^{\rm mis} + j_{b}^{\rm mis}  j_{b}^{\rm mis},
\end{align}
where $ j_{b}^{\rm mis}$ is the light jet (from $u$, $d$, $s$, $c$, and $g$) mistagged as a $b$ quark jet.
We calculated the parton-level cross sections of the backgrounds,
with the $b$ tagging and mistagging efficiencies of $P_{b\to b} =0.7$, $P_{c\to b}=0.05$, and $P_{j\to b}=0.01$.
We imposed the selection cuts of $p_T^b>20\gev$, $p_T^\ell>10\gev$, $|\eta^{\ell,j}|<2.5$, 
$\met>20\gev$, and the separation $\Delta R_{ii'}>0.4$.
After the basic selection, the background cross section from $\ttop jj$ is about $ 8.7\ab$
and from $\ttop \ell\nu$ is about $3.8\times 10^{-4}\ab$.
If we impose additional cuts on the invariant mass of two $b$ jets like $|m_{bb}-\mh|<15\gev$,
the backgrounds are negligible.
Despite the almost background-free environment,
the high $\lmc$ yields too small signal rate of $pp\to \ch h h$ at the 14 TeV LHC.

%
%
%

Exploring the cutoff scale in the inverted type I via the $6b+\ell\nu$ final state
has a better chance in future high-energy colliders 
such as  the Future hadron-hadron Circular Collider (FCC-hh) at CERN~\cite{Gomez-Ceballos:2013zzn},
the CEPC~\cite{Gao:2021bam, CEPCStudyGroup:2018ghi}, and the muon collider~\cite{Palmer:2007zzc,Delahaye:2019omf,Black:2022cth}.
Particularly, we have high expectations for the muon collider with benchmark energies in the range of
$\sqrt{s}=3-30\tev$ and the integrated luminosity of 
$\mathcal{L} = 10 \,(\sqrt{s}/10\tev)^2 \iab$.
The triple Higgs processes in \eq{eq:LHC:Ahh} and \eq{eq:LHC:hhh}
with $\qq$ replaced by $\mmu$
will be able to disentangle the high- and low-cutoff scales in the inverted type I.

\section{Conclusion}
\label{sec:conclusions}
Beyond the studies on how high the cutoff scale of a new physics model can go up,
we have pursued an efficient observable to distinguish the high- and low-cutoff scales.
The type I in the 2HDM has been considered for the inverted scenario where
the observed Higgs boson at a mass of $125\gev$ is the heavier \textit{CP}-even Higgs boson $H$.
We have first obtained the still-available parameter points
that satisfy the theoretical requirements, the experimental constraints, and the cutoff scale above $1\tev$.
The viable parameter space at the electroweak scale is already limited 
such that $\ma,\mch\lsim 430\gev$ and $\mh \gsim 62.5\gev$.
Through the calculation of the cutoff scale $\lmc$ of each viable parameter point by using the RGE,
we have shown that the inverted type I can retain the stability all the way up to the Planck scale.

The condition of $\lmc>10^{18}\gev$ requires the light masses of the extra Higgs bosons 
like $\ma, \mch\lsim 160\gev$.
However, the light masses alone cannot guarantee the high-cutoff scale
because the parameter points with light masses accommodate $\lmc$ from $1\tev$ to $10^{19}\gev$.
Targeting at the phenomenologically challenging case of $\tb>10$,
we have investigated the trilinear Higgs couplings versus $\lmc$.
Although the values of all the trilinear couplings for $\lmc=1\tev$
are different from those for $\lmc=10^{18}\gev$,
$\lmh_{hhh}$ shows a large variation about $\lmc$, 
$\lmh_{hhh} \in [-0.09,\, 1.1]$.
Multi-Higgs boson productions at the LHC have been studied 
to probe the cutoff scale.
The gluon fusion productions of the di-Higgs, $gg\to hh/AA$,
are not efficient to measure $\lmh_{hhh}$ because the dominant contribution from $H$ dilutes the $h$ contribution.
The most remarkable result is in the tri-Higgs process of $pp \to \ch hh$ mediated by the $W$ boson.
The signal cross section shows huge variation according to $\lmc$,
like $\sg_{\lmc=1\tev}/\sg_{\lmc=10^{18}\gev} \sim 10^3$.
The precision measurement of $pp \to \ch hh$
can indeed distinguish the high- and low-cutoff scales of the model.
Considering the dominant decay modes of $\ch \to W^{\pm *}A/h$, $h \to bb$, and $A\to bb$,
an efficient final state of $pp \to \ch hh$ is $6b +\ell\nu$.
Although it enjoys almost background-free environment at the LHC,
the small cross section ($\lsim 1\ab$) for $\lmc=10^{18}\gev$ 
motivates future high-energy colliders, especially the muon collider with $\sqrt{s}=3-30\tev$.

\acknowledgments
The work of J.~K., S.~L, and J.~S is supported by 
the National Research Foundation of Korea, Grant No.~NRF-2022R1A2C1007583. 
S.~K.~K. was supported by the National Research Foundation of Korea (NRF) grant funded by the Korea government (MSIT) (No.2019R1A2C1088953).

\appendix

\section{RGEs in the type I}
\label{appendix:RGE}
Focusing on the type I, we present the one-loop level RGEs~\cite{Cheng:1973nv, Komatsu:1981xh, Branco:2011iw, Das:2015mwa, Basler:2017nzu}.
The beta functions of gauge couplings are given by
\begin{align}
\label{eq:RGE:g}
16\pi^{2} \beta_{g_3} &= -7 g_3^3, \\
16\pi^{2} \beta_{g_2} &= \left(-\frac{10}{3}+\frac{n_{d} }{6}\right) g_2^3 =-3 g_2^3, \\
16\pi^{2} \beta_{g_1} &= \left(\frac{20}{3}+\frac{n_{d}}{6}\right) = 7 g_1^3,
\end{align}
where $n_d$ is the number of the scalar doublets of the fermions, so $n_{d}=2$ in the 2HDM. 
The running of the quartic couplings of $\lambda_{i}$'s is different according to the type.
First, we write the $\beta$ functions  
in terms of the common part $c_i$ and the type-dependent part $h_i$ as
\begin{align}
\label{eq:RGE:lm}
16\pi^{2}\beta_{\lambda_{i}} ^\tp
&= c_i + h_i^\tp, \quad (i=1,\cdots,5).
\end{align}
The common parts for $\lm_i$'s are
\begin{align}
\label{eq:ci}
c_1
&=
12\lambda_{1}^{2}+4\lambda_{3}^{2}+4\lambda_{3}\lambda_{4}+2\lambda_{4}^{2}+2\lambda_{5}^{2}
+\frac{3}{4}(3g^{4}+g'^{4}+2g^{2}g'^{2})-3\lambda_{1}(3g^{2}+g'^{2}), \\[3pt] \nn
c_2
&=
12\lambda_{2}^{2}+4\lambda_{3}^{2}+4\lambda_{3}\lambda_{4}+2\lambda_{4}^{2}+2\lambda_{5}^{2}
+\frac{3}{4}(3g^{4}+g'^{4}+2g^{2}g'^{2})-3\lambda_{2}(3g^{2}+g'^{2})  \\[3pt] \nn
&\quad
+ 12 y_{t}^{2} \lambda_{2} - 12 y_{t}^{4},
\\[3pt] \nn
c_3
&=
(\lambda_{1}+\lambda_{2})(6\lambda_{3}+2\lambda_{4})+4\lambda_{3}^{2}+2\lambda_{4}^{2}+2\lambda_{5}^{2}
+\frac{3}{4}(3g^{4}+g'^{4}-2g^{2}g'^{2})-3\lambda_{3}(3g^{2}+g'^{2})  \\[3pt] \nn
&\quad
+2(3y_{t}^{2}+3y_{b}^{2}+y_{\tau}^{2})\lambda_{3}, \\[3pt] \nn
c_4
&=
2(\lambda_{1}+\lambda_{2})\lambda_{4}+8\lambda_{3}\lambda_{4}+4\lambda_{4}^{2}+8\lambda_{5}^{2}
+3g^{2}g'^{2}-3(3g^{2}+g'^{2})\lambda_{4}  \\[3pt] \nn
&\quad
+2(3y_{t}^{2}+3y_{b}^{2}+y_{\tau}^{2})\lambda_{4}, \\[3pt] \nn
c_5
&=
2(\lambda_{1}+\lambda_{2}+4\lambda_{3}+6\lambda_{4})\lambda_{5}-3\lambda_{5}(3g^{2}+g'^{2})
+2(3y_{t}^{2}+3y_{b}^{2}+y_{\tau}^{2})\lambda_{5}.
\end{align}
The $h_{i}$'s in type I are 
\begin{align}
h_1^{\rm I} &=  h_3^{\rm I} =  h_4^{\rm I} =h_5^{\rm I}=0,\\ \nn
h_2^{\rm I} &= 4(3y_{b}^{2}+y_{\tau}^{2})\lambda_{2}-4(3y_{b}^{4}+y_{\tau}^{4}).
\end{align}

The Yukawa couplings of the top quark, bottom quark, and tau lepton ($y_t$, $y_b$, and $y_\tau$) 
are running with the $\beta$ functions of
\begin{align}
\label{eq:RGE:y}
16\pi^{2}\beta_{y_f} ^\tp
&= c_{y_f}+ h_{y_f}^\tp, \quad (f=t,b,\tau),
\end{align}
where the common parts are
\begin{align}
\label{eq:cyf}
c_{y_{t}}
&=\left(-8g_{s}^{2}-\frac{9}{4}g^{2}-\frac{17}{12}g'^{2}+\frac{9}{2}y_{t}^{2}\right)y_{t}, \\[3pt] 
c_{y_{b}}
&=\left(-8g_{s}^{2}-\frac{9}{4}g^{2}-\frac{5}{12}g'^{2}+\frac{3}{2}y_{t}^{2}+\frac{9}{2}y_{b}^{2}\right)y_{b}, \\[3pt] \nn
c_{y_{\tau}}
&=\left(-\frac{9}{4}g^{2}-\frac{15}{4}g'^{2}+\frac{5}{2}y_{\tau}^{2}\right)y_{\tau},
\end{align}
and the type-dependent parts are 
\begin{equation}
\label{eq:hyf}
h_{y_t}^{\rm I} = \left(\frac{3}{2}y_{b}^{2}+y_{\tau}^{2}\right)y_{t},
\quad
h_{y_b}^{\rm I} =y_\tau^2 y_b,
\quad
h_{y_\tau}^{\rm I} =3 \lf y_{t}^{2}+ y_{b}^{2}\ri y_\tau.
\end{equation}
The initial conditions of the Yukawa coupling  are set at the top quark mass scale $m_t^{\rm pole}$~\cite{Das:2015mwa} as
\begin{align}
y_t(m_t^{\rm pole}) &= \frac{\sqrt{2} m_t}{v \sb} \left\{ 1-\frac{4 }{3\pi} \alpha_s(m_t) \right\},
\\ \nn
y_b(m_t^{\rm pole}) &=\frac{\sqrt{2} m_b}{v \sb},
\\ \nn
y_\tau(m_t^{\rm pole}) &= \frac{\sqrt{2} m_\tau}{v \sb} .
\end{align}


\begin{thebibliography}{100}

\bibitem{Ellis:2021kzk}
J.~Ellis, \textit{{SMEFT Constraints on New Physics beyond the Standard
  Model}},  in \emph{{Beyond Standard Model: From Theory to Experiment}}, 5,
  2021.
\newblock \href{https://arxiv.org/abs/2105.14942}{\texttt{2105.14942}}.
\newblock \href{http://dx.doi.org/10.31526/ACP.BSM-2021.16}{DOI}.

\bibitem{ATLAS:2020fcp}
{\scshape ATLAS} collaboration, G.~Aad et~al., \textit{{Measurements of $WH$
  and $ZH$ production in the $H \rightarrow b\bar{b}$ decay channel in $pp$
  collisions at 13 TeV with the ATLAS detector}},
  \href{http://dx.doi.org/10.1140/epjc/s10052-020-08677-2}{\emph{Eur. Phys. J.
  C} \textbf{ 81} (2021) 178},
  [\href{https://arxiv.org/abs/2007.02873}{\texttt{2007.02873}}].

\bibitem{ATLAS:2020bhl}
{\scshape ATLAS} collaboration, G.~Aad et~al., \textit{{Measurements of Higgs
  bosons decaying to bottom quarks from vector boson fusion production with the
  ATLAS experiment at $\sqrt{s}=13\,\text {TeV}$}},
  \href{http://dx.doi.org/10.1140/epjc/s10052-021-09192-8}{\emph{Eur. Phys. J.
  C} \textbf{ 81} (2021) 537},
  [\href{https://arxiv.org/abs/2011.08280}{\texttt{2011.08280}}].

\bibitem{CMS:2020zge}
{\scshape CMS} collaboration, A.~M. Sirunyan et~al., \textit{{Inclusive search
  for highly boosted Higgs bosons decaying to bottom quark-antiquark pairs in
  proton-proton collisions at $\sqrt{s} =$ 13 TeV}},
  \href{http://dx.doi.org/10.1007/JHEP12(2020)085}{\emph{JHEP} \textbf{ 12}
  (2020) 085}, [\href{https://arxiv.org/abs/2006.13251}{\texttt{2006.13251}}].

\bibitem{ATLAS:2021nsx}
{\scshape ATLAS} collaboration, \textit{{Study of Higgs-boson production with
  large transverse momentum using the $H\rightarrow b\bar{b}$ decay with the
  ATLAS detector}}, [\href{http://cds.cern.ch/record/2759284}{\texttt{http://cds.cern.ch/record/2759284}}].

\bibitem{CMS:2021gxc}
{\scshape CMS} collaboration, A.~Tumasyan et~al., \textit{{Measurement of the
  inclusive and differential Higgs boson production cross sections in the decay
  mode to a pair of $\tau$ leptons in pp collisions at $\sqrt{s} = $ 13 TeV}},
  \href{http://dx.doi.org/10.1103/PhysRevLett.128.081805}{\emph{Phys. Rev.
  Lett.} \textbf{ 128} (2022) 081805},
  [\href{https://arxiv.org/abs/2107.11486}{\texttt{2107.11486}}].

\bibitem{ATLAS:2020syy}
{\scshape ATLAS} collaboration, \textit{{Measurement of the Higgs boson
  decaying to $b$-quarks produced in association with a top-quark pair in $pp$
  collisions at $\sqrt{s}=13$ TeV with the ATLAS detector}}, .

\bibitem{ATLAS:2021upe}
{\scshape ATLAS} collaboration, \textit{{Measurements of gluon fusion and
  vector-boson-fusion production of the Higgs boson in $H\rightarrow W W^*
  \rightarrow e\nu \mu\nu$ decays using $pp$ collisions at $\sqrt{s}=13$ TeV
  with the ATLAS detector}},  [\href{http://cds.cern.ch/record/2743685}{\texttt{http://cds.cern.ch/record/2743685}}].


\bibitem{ATLAS:2020pvn}
{\scshape ATLAS} collaboration, \textit{{Measurement of the properties of Higgs
  boson production at $\sqrt{s}$=13 TeV in the $H\to \gamma\gamma$ channel
  using 139 fb$^{−1}$ of $pp$ collision data with the ATLAS experiment}}, 
  [\href{http://cds.cern.ch/record/2725727}{\texttt{http://cds.cern.ch/record/2725727}}].
  
\bibitem{CMS:2021ugl}
{\scshape CMS} collaboration, A.~M. Sirunyan et~al., \textit{{Measurements of
  production cross sections of the Higgs boson in the four-lepton final state
  in proton\textendash{}proton collisions at $\sqrt{s} = 13\,\text {Te}\text
  {V} $}}, \href{http://dx.doi.org/10.1140/epjc/s10052-021-09200-x}{\emph{Eur.
  Phys. J. C} \textbf{ 81} (2021) 488},
  [\href{https://arxiv.org/abs/2103.04956}{\texttt{2103.04956}}].

\bibitem{ATLAS:2020wny}
{\scshape ATLAS} collaboration, G.~Aad et~al., \textit{{Measurements of the
  Higgs boson inclusive and differential fiducial cross sections in the 4$\ell$
  decay channel at $\sqrt{s}$ = 13 TeV}},
  \href{http://dx.doi.org/10.1140/epjc/s10052-020-8223-0}{\emph{Eur. Phys. J.
  C} \textbf{ 80} (2020) 942},
  [\href{https://arxiv.org/abs/2004.03969}{\texttt{2004.03969}}].

\bibitem{ATLAS:2020rej}
{\scshape ATLAS} collaboration, G.~Aad et~al., \textit{{Higgs boson production
  cross-section measurements and their EFT interpretation in the $4\ell $ decay
  channel at $\sqrt{s}=$13 TeV with the ATLAS detector}},
  \href{http://dx.doi.org/10.1140/epjc/s10052-020-8227-9}{\emph{Eur. Phys. J.
  C} \textbf{ 80} (2020) 957},
  [\href{https://arxiv.org/abs/2004.03447}{\texttt{2004.03447}}].

\bibitem{ATLAS:2020qdt}
{\scshape ATLAS} collaboration, \textit{{A combination of measurements of Higgs
  boson production and decay using up to $139$ fb$^{-1}$ of proton--proton
  collision data at $\sqrt{s}=$ 13 TeV collected with the ATLAS experiment}}, 
  [\href{http://cds.cern.ch/record/2725733}{\texttt{http://cds.cern.ch/record/2725733}}].
  
\bibitem{ATLAS:2020fzp}
{\scshape ATLAS} collaboration, G.~Aad et~al., \textit{{A search for the dimuon
  decay of the Standard Model Higgs boson with the ATLAS detector}},
  \href{http://dx.doi.org/10.1016/j.physletb.2020.135980}{\emph{Phys. Lett. B}
  \textbf{ 812} (2021) 135980},
  [\href{https://arxiv.org/abs/2007.07830}{\texttt{2007.07830}}].

\bibitem{CMS:2020xwi}
{\scshape CMS} collaboration, A.~M. Sirunyan et~al., \textit{{Evidence for
  Higgs boson decay to a pair of muons}},
  \href{http://dx.doi.org/10.1007/JHEP01(2021)148}{\emph{JHEP} \textbf{ 01}
  (2021) 148}, [\href{https://arxiv.org/abs/2009.04363}{\texttt{2009.04363}}].

\bibitem{ATLAS:2021zwx}
{\scshape ATLAS} collaboration, \textit{{Direct constraint on the Higgs-charm
  coupling from a search for Higgs boson decays to charm quarks with the ATLAS
  detector}}, [\href{http://cds.cern.ch/record/2771724}{\texttt{http://cds.cern.ch/record/2771724}}].

\bibitem{Turok:1990zg}
N.~Turok and J.~Zadrozny, \textit{{Electroweak baryogenesis in the two doublet
  model}}, \href{http://dx.doi.org/10.1016/0550-3213(91)90356-3}{\emph{Nucl.
  Phys. B} \textbf{ 358} (1991) 471--493}.

\bibitem{Cohen:1991iu}
A.~G. Cohen, D.~B. Kaplan and A.~E. Nelson, \textit{{Spontaneous baryogenesis
  at the weak phase transition}},
  \href{http://dx.doi.org/10.1016/0370-2693(91)91711-4}{\emph{Phys. Lett. B}
  \textbf{ 263} (1991) 86--92}.

\bibitem{Zarikas:1995qb}
V.~Zarikas, \textit{{The Phase transition of the two Higgs extension of the
  standard model}},
  \href{http://dx.doi.org/10.1016/0370-2693(96)00701-0}{\emph{Phys. Lett. B}
  \textbf{ 384} (1996) 180--184},
  [\href{https://arxiv.org/abs/hep-ph/9509338}{\texttt{hep-ph/9509338}}].

\bibitem{Cline:1996mga}
J.~M. Cline and P.-A. Lemieux, \textit{{Electroweak phase transition in two
  Higgs doublet models}},
  \href{http://dx.doi.org/10.1103/PhysRevD.55.3873}{\emph{Phys. Rev. D}
  \textbf{ 55} (1997) 3873--3881},
  [\href{https://arxiv.org/abs/hep-ph/9609240}{\texttt{hep-ph/9609240}}].

\bibitem{Fromme:2006cm}
L.~Fromme, S.~J. Huber and M.~Seniuch, \textit{{Baryogenesis in the two-Higgs
  doublet model}},
  \href{http://dx.doi.org/10.1088/1126-6708/2006/11/038}{\emph{JHEP} \textbf{
  11} (2006) 038},
  [\href{https://arxiv.org/abs/hep-ph/0605242}{\texttt{hep-ph/0605242}}].

\bibitem{Machacek:1983tz}
M.~E. Machacek and M.~T. Vaughn, \textit{{Two Loop Renormalization Group
  Equations in a General Quantum Field Theory. 1. Wave Function
  Renormalization}},
  \href{http://dx.doi.org/10.1016/0550-3213(83)90610-7}{\emph{Nucl. Phys. B}
  \textbf{ 222} (1983) 83--103}.

\bibitem{Machacek:1983fi}
M.~E. Machacek and M.~T. Vaughn, \textit{{Two Loop Renormalization Group
  Equations in a General Quantum Field Theory. 2. Yukawa Couplings}},
  \href{http://dx.doi.org/10.1016/0550-3213(84)90533-9}{\emph{Nucl. Phys. B}
  \textbf{ 236} (1984) 221--232}.

\bibitem{Machacek:1984zw}
M.~E. Machacek and M.~T. Vaughn, \textit{{Two Loop Renormalization Group
  Equations in a General Quantum Field Theory. 3. Scalar Quartic Couplings}},
  \href{http://dx.doi.org/10.1016/0550-3213(85)90040-9}{\emph{Nucl. Phys. B}
  \textbf{ 249} (1985) 70--92}.

\bibitem{Luo:2002ti}
M.-x. Luo, H.-w. Wang and Y.~Xiao, \textit{{Two loop renormalization group
  equations in general gauge field theories}},
  \href{http://dx.doi.org/10.1103/PhysRevD.67.065019}{\emph{Phys. Rev. D}
  \textbf{ 67} (2003) 065019},
  [\href{https://arxiv.org/abs/hep-ph/0211440}{\texttt{hep-ph/0211440}}].

\bibitem{Das:2015mwa}
D.~Das and I.~Saha, \textit{{Search for a stable alignment limit in
  two-Higgs-doublet models}},
  \href{http://dx.doi.org/10.1103/PhysRevD.91.095024}{\emph{Phys. Rev. D}
  \textbf{ 91} (2015) 095024},
  [\href{https://arxiv.org/abs/1503.02135}{\texttt{1503.02135}}].

\bibitem{Haber:1993an}
H.~E. Haber and R.~Hempfling, \textit{{The Renormalization group improved Higgs
  sector of the minimal supersymmetric model}},
  \href{http://dx.doi.org/10.1103/PhysRevD.48.4280}{\emph{Phys. Rev. D}
  \textbf{ 48} (1993) 4280--4309},
  [\href{https://arxiv.org/abs/hep-ph/9307201}{\texttt{hep-ph/9307201}}].

\bibitem{Grimus:2004yh}
W.~Grimus and L.~Lavoura, \textit{{Renormalization of the neutrino mass
  operators in the multi-Higgs-doublet standard model}},
  \href{http://dx.doi.org/10.1140/epjc/s2004-02075-0}{\emph{Eur. Phys. J. C}
  \textbf{ 39} (2005) 219--227},
  [\href{https://arxiv.org/abs/hep-ph/0409231}{\texttt{hep-ph/0409231}}].

\bibitem{Chowdhury:2015yja}
D.~Chowdhury and O.~Eberhardt, \textit{{Global fits of the two-loop
  renormalized Two-Higgs-Doublet model with soft Z$_{2}$ breaking}},
  \href{http://dx.doi.org/10.1007/JHEP11(2015)052}{\emph{JHEP} \textbf{ 11}
  (2015) 052}, [\href{https://arxiv.org/abs/1503.08216}{\texttt{1503.08216}}].

\bibitem{Basler:2017nzu}
P.~Basler, P.~M. Ferreira, M.~M\"uhlleitner and R.~Santos, \textit{{High scale
  impact in alignment and decoupling in two-Higgs doublet models}},
  \href{http://dx.doi.org/10.1103/PhysRevD.97.095024}{\emph{Phys. Rev. D}
  \textbf{ 97} (2018) 095024},
  [\href{https://arxiv.org/abs/1710.10410}{\texttt{1710.10410}}].

\bibitem{Krauss:2018thf}
M.~E. Krauss, T.~Opferkuch and F.~Staub, \textit{{The Ultraviolet Landscape of
  Two-Higgs Doublet Models}},
  \href{http://dx.doi.org/10.1140/epjc/s10052-018-6489-2}{\emph{Eur. Phys. J.
  C} \textbf{ 78} (2018) 1020},
  [\href{https://arxiv.org/abs/1807.07581}{\texttt{1807.07581}}].

\bibitem{Oredsson:2018yho}
J.~Oredsson and J.~Rathsman, \textit{{$\mathbb Z_2$ breaking effects in 2-loop
  RG evolution of 2HDM}},
  \href{http://dx.doi.org/10.1007/JHEP02(2019)152}{\emph{JHEP} \textbf{ 02}
  (2019) 152}, [\href{https://arxiv.org/abs/1810.02588}{\texttt{1810.02588}}].

\bibitem{Aiko:2020atr}
M.~Aiko and S.~Kanemura, \textit{{New scenario for aligned Higgs couplings
  originated from the twisted custodial symmetry at high energies}},
  \href{http://dx.doi.org/10.1007/JHEP02(2021)046}{\emph{JHEP} \textbf{ 02}
  (2021) 046}, [\href{https://arxiv.org/abs/2009.04330}{\texttt{2009.04330}}].

\bibitem{Dey:2021pyn}
A.~Dey, J.~Lahiri and B.~Mukhopadhyaya, \textit{{Muon g-2 and a type-X
  two-Higgs-doublet scenario: Some studies in high-scale validity}},
  \href{http://dx.doi.org/10.1103/PhysRevD.106.055023}{\emph{Phys. Rev. D}
  \textbf{ 106} (2022) 055023},
  [\href{https://arxiv.org/abs/2106.01449}{\texttt{2106.01449}}].

\bibitem{Lee:2022gyf}
S.~Lee, K.~Cheung, J.~Kim, C.-T. Lu and J.~Song, \textit{{Status of the
  two-Higgs-doublet model in light of the CDF mW measurement}},
  \href{http://dx.doi.org/10.1103/PhysRevD.106.075013}{\emph{Phys. Rev. D}
  \textbf{ 106} (2022) 075013},
  [\href{https://arxiv.org/abs/2204.10338}{\texttt{2204.10338}}].

\bibitem{Kim:2022hvh}
J.~Kim, S.~Lee, P.~Sanyal and J.~Song, \textit{{CDF W-boson mass and muon g-2
  in a type-X two-Higgs-doublet model with a Higgs-phobic light pseudoscalar}},
  \href{http://dx.doi.org/10.1103/PhysRevD.106.035002}{\emph{Phys. Rev. D}
  \textbf{ 106} (2022) 035002},
  [\href{https://arxiv.org/abs/2205.01701}{\texttt{2205.01701}}].

\bibitem{Kim:2022nmm}
J.~Kim, S.~Lee, J.~Song and P.~Sanyal, \textit{{Fermiophobic light Higgs boson
  in the type-I two-Higgs-doublet model}},
  \href{http://dx.doi.org/10.1016/j.physletb.2022.137406}{\emph{Phys. Lett. B}
  \textbf{ 834} (2022) 137406},
  [\href{https://arxiv.org/abs/2207.05104}{\texttt{2207.05104}}].

\bibitem{Ferreira:2012my}
P.~M. Ferreira, R.~Santos, M.~Sher and J.~P. Silva, \textit{{Could the LHC
  two-photon signal correspond to the heavier scalar in two-Higgs-doublet
  models?}}, \href{http://dx.doi.org/10.1103/PhysRevD.85.035020}{\emph{Phys.
  Rev. D} \textbf{ 85} (2012) 035020},
  [\href{https://arxiv.org/abs/1201.0019}{\texttt{1201.0019}}].

\bibitem{Chang:2015goa}
S.~Chang, S.~K. Kang, J.-P. Lee and J.~Song, \textit{{Higgs potential and
  hidden light Higgs scenario in two Higgs doublet models}},
  \href{http://dx.doi.org/10.1103/PhysRevD.92.075023}{\emph{Phys. Rev. D}
  \textbf{ 92} (2015) 075023},
  [\href{https://arxiv.org/abs/1507.03618}{\texttt{1507.03618}}].

\bibitem{Jueid:2021avn}
A.~Jueid, J.~Kim, S.~Lee and J.~Song, \textit{{Type-X two-Higgs-doublet model
  in light of the muon g-2: Confronting Higgs boson and collider data}},
  \href{http://dx.doi.org/10.1103/PhysRevD.104.095008}{\emph{Phys. Rev. D}
  \textbf{ 104} (2021) 095008},
  [\href{https://arxiv.org/abs/2104.10175}{\texttt{2104.10175}}].

\bibitem{Branco:2011iw}
G.~C. Branco, P.~M. Ferreira, L.~Lavoura, M.~N. Rebelo, M.~Sher and J.~P.
  Silva, \textit{{Theory and phenomenology of two-Higgs-doublet models}},
  \href{http://dx.doi.org/10.1016/j.physrep.2012.02.002}{\emph{Phys. Rept.}
  \textbf{ 516} (2012) 1--102},
  [\href{https://arxiv.org/abs/1106.0034}{\texttt{1106.0034}}].

\bibitem{Glashow:1976nt}
S.~L. Glashow and S.~Weinberg, \textit{{Natural Conservation Laws for Neutral
  Currents}}, \href{http://dx.doi.org/10.1103/PhysRevD.15.1958}{\emph{Phys.
  Rev. D} \textbf{ 15} (1977) 1958}.

\bibitem{Paschos:1976ay}
E.~A. Paschos, \textit{{Diagonal Neutral Currents}},
  \href{http://dx.doi.org/10.1103/PhysRevD.15.1966}{\emph{Phys. Rev. D}
  \textbf{ 15} (1977) 1966}.

\bibitem{Haber:2006ue}
H.~E. Haber and D.~O'Neil, \textit{{Basis-independent methods for the
  two-Higgs-doublet model. II. The Significance of tan$\beta$}},
  \href{http://dx.doi.org/10.1103/PhysRevD.74.015018}{\emph{Phys. Rev. D}
  \textbf{ 74} (2006) 015018},
  [\href{https://arxiv.org/abs/hep-ph/0602242}{\texttt{hep-ph/0602242}}].

\bibitem{Song:2019aav}
J.~Song and Y.~W. Yoon, \textit{{$W\gamma$ decay of the elusive charged Higgs
  boson in the two-Higgs-doublet model with vectorlike fermions}},
  \href{http://dx.doi.org/10.1103/PhysRevD.100.055006}{\emph{Phys. Rev. D}
  \textbf{ 100} (2019) 055006},
  [\href{https://arxiv.org/abs/1904.06521}{\texttt{1904.06521}}].

\bibitem{Kanemura:2011sj}
S.~Kanemura, Y.~Okada, H.~Taniguchi and K.~Tsumura, \textit{{Indirect bounds on
  heavy scalar masses of the two-Higgs-doublet model in light of recent Higgs
  boson searches}},
  \href{http://dx.doi.org/10.1016/j.physletb.2011.09.035}{\emph{Phys. Lett. B}
  \textbf{ 704} (2011) 303--307},
  [\href{https://arxiv.org/abs/1108.3297}{\texttt{1108.3297}}].

\bibitem{Bernon:2015qea}
J.~Bernon, J.~F. Gunion, H.~E. Haber, Y.~Jiang and S.~Kraml,
  \textit{{Scrutinizing the alignment limit in two-Higgs-doublet models:
  m$_h$=125 GeV}},
  \href{http://dx.doi.org/10.1103/PhysRevD.92.075004}{\emph{Phys. Rev. D}
  \textbf{ 92} (2015) 075004},
  [\href{https://arxiv.org/abs/1507.00933}{\texttt{1507.00933}}].

\bibitem{Bernon:2015wef}
J.~Bernon, J.~F. Gunion, H.~E. Haber, Y.~Jiang and S.~Kraml,
  \textit{{Scrutinizing the alignment limit in two-Higgs-doublet models. II.
  m$_H$=125 GeV}},
  \href{http://dx.doi.org/10.1103/PhysRevD.93.035027}{\emph{Phys. Rev. D}
  \textbf{ 93} (2016) 035027},
  [\href{https://arxiv.org/abs/1511.03682}{\texttt{1511.03682}}].

\bibitem{Plehn:1996wb}
T.~Plehn, M.~Spira and P.~M. Zerwas, \textit{{Pair production of neutral Higgs
  particles in gluon-gluon collisions}},
  \href{http://dx.doi.org/10.1016/0550-3213(96)00418-X}{\emph{Nucl. Phys. B}
  \textbf{ 479} (1996) 46--64},
  [\href{https://arxiv.org/abs/hep-ph/9603205}{\texttt{hep-ph/9603205}}].

\bibitem{Djouadi:1999rca}
A.~Djouadi, W.~Kilian, M.~Muhlleitner and P.~M. Zerwas, \textit{{Production of
  neutral Higgs boson pairs at LHC}},
  \href{http://dx.doi.org/10.1007/s100529900083}{\emph{Eur. Phys. J. C}
  \textbf{ 10} (1999) 45--49},
  [\href{https://arxiv.org/abs/hep-ph/9904287}{\texttt{hep-ph/9904287}}].

\bibitem{Barger:2013jfa}
V.~Barger, L.~L. Everett, C.~B. Jackson and G.~Shaughnessy, \textit{{Higgs-Pair
  Production and Measurement of the Triscalar Coupling at LHC(8,14)}},
  \href{http://dx.doi.org/10.1016/j.physletb.2013.12.013}{\emph{Phys. Lett. B}
  \textbf{ 728} (2014) 433--436},
  [\href{https://arxiv.org/abs/1311.2931}{\texttt{1311.2931}}].

\bibitem{Dawson:2015oha}
S.~Dawson, A.~Ismail and I.~Low, \textit{{What\textquoteright{}s in the loop?
  The anatomy of double Higgs production}},
  \href{http://dx.doi.org/10.1103/PhysRevD.91.115008}{\emph{Phys. Rev. D}
  \textbf{ 91} (2015) 115008},
  [\href{https://arxiv.org/abs/1504.05596}{\texttt{1504.05596}}].

\bibitem{Cheung:2020xij}
K.~Cheung, A.~Jueid, C.-T. Lu, J.~Song and Y.~W. Yoon, \textit{{Disentangling
  new physics effects on nonresonant Higgs boson pair production from gluon
  fusion}}, \href{http://dx.doi.org/10.1103/PhysRevD.103.015019}{\emph{Phys.
  Rev. D} \textbf{ 103} (2021) 015019},
  [\href{https://arxiv.org/abs/2003.11043}{\texttt{2003.11043}}].

\bibitem{Jueid:2021qfs}
A.~Jueid, J.~Kim, S.~Lee and J.~Song, \textit{{Studies of nonresonant Higgs
  pair production at electron-proton colliders}},
  \href{http://dx.doi.org/10.1016/j.physletb.2021.136417}{\emph{Phys. Lett. B}
  \textbf{ 819} (2021) 136417},
  [\href{https://arxiv.org/abs/2102.12507}{\texttt{2102.12507}}].

\bibitem{Ivanov:2006yq}
I.~P. Ivanov, \textit{{Minkowski space structure of the Higgs potential in
  2HDM}}, \href{http://dx.doi.org/10.1103/PhysRevD.75.035001}{\emph{Phys. Rev.
  D} \textbf{ 75} (2007) 035001},
  [\href{https://arxiv.org/abs/hep-ph/0609018}{\texttt{hep-ph/0609018}}].

\bibitem{Ginzburg:2005dt}
I.~F. Ginzburg and I.~P. Ivanov, \textit{{Tree-level unitarity constraints in
  the most general 2HDM}},
  \href{http://dx.doi.org/10.1103/PhysRevD.72.115010}{\emph{Phys. Rev. D}
  \textbf{ 72} (2005) 115010},
  [\href{https://arxiv.org/abs/hep-ph/0508020}{\texttt{hep-ph/0508020}}].

\bibitem{Kanemura:2015ska}
S.~Kanemura and K.~Yagyu, \textit{{Unitarity bound in the most general two
  Higgs doublet model}},
  \href{http://dx.doi.org/10.1016/j.physletb.2015.10.047}{\emph{Phys. Lett. B}
  \textbf{ 751} (2015) 289--296},
  [\href{https://arxiv.org/abs/1509.06060}{\texttt{1509.06060}}].

\bibitem{Arhrib:2000is}
A.~Arhrib, \textit{{Unitarity constraints on scalar parameters of the standard
  and two Higgs doublets model}},  in \emph{{Workshop on Noncommutative
  Geometry, Superstrings and Particle Physics}}, 12, 2000.
\newblock \href{https://arxiv.org/abs/hep-ph/0012353}{\texttt{hep-ph/0012353}}.

\bibitem{Ivanov:2008cxa}
I.~P. Ivanov, \textit{{General two-order-parameter Ginzburg-Landau model with
  quadratic and quartic interactions}},
  \href{http://dx.doi.org/10.1103/PhysRevE.79.021116}{\emph{Phys. Rev. E}
  \textbf{ 79} (2009) 021116},
  [\href{https://arxiv.org/abs/0802.2107}{\texttt{0802.2107}}].

\bibitem{Barroso:2012mj}
A.~Barroso, P.~M. Ferreira, I.~P. Ivanov, R.~Santos and J.~P. Silva,
  \textit{{Evading death by vacuum}},
  \href{http://dx.doi.org/10.1140/epjc/s10052-013-2537-0}{\emph{Eur. Phys. J.
  C} \textbf{ 73} (2013) 2537},
  [\href{https://arxiv.org/abs/1211.6119}{\texttt{1211.6119}}].

\bibitem{Barroso:2013awa}
A.~Barroso, P.~M. Ferreira, I.~P. Ivanov and R.~Santos, \textit{{Metastability
  bounds on the two Higgs doublet model}},
  \href{http://dx.doi.org/10.1007/JHEP06(2013)045}{\emph{JHEP} \textbf{ 06}
  (2013) 045}, [\href{https://arxiv.org/abs/1303.5098}{\texttt{1303.5098}}].

\bibitem{Eriksson:2009ws}
D.~Eriksson, J.~Rathsman and O.~Stal, \textit{{2HDMC: Two-Higgs-Doublet Model
  Calculator Physics and Manual}},
  \href{http://dx.doi.org/10.1016/j.cpc.2009.09.011}{\emph{Comput. Phys.
  Commun.} \textbf{ 181} (2010) 189--205},
  [\href{https://arxiv.org/abs/0902.0851}{\texttt{0902.0851}}].

\bibitem{Branchina:2018qlf}
V.~Branchina, F.~Contino and P.~M. Ferreira, \textit{{Electroweak vacuum
  lifetime in two Higgs doublet models}},
  \href{http://dx.doi.org/10.1007/JHEP11(2018)107}{\emph{JHEP} \textbf{ 11}
  (2018) 107}, [\href{https://arxiv.org/abs/1807.10802}{\texttt{1807.10802}}].

\bibitem{Peskin:1991sw}
M.~E. Peskin and T.~Takeuchi, \textit{{Estimation of oblique electroweak
  corrections}}, \href{http://dx.doi.org/10.1103/PhysRevD.46.381}{\emph{Phys.
  Rev. D} \textbf{ 46} (1992) 381--409}.

\bibitem{He:2001tp}
H.-J. He, N.~Polonsky and S.-f. Su, \textit{{Extra families, Higgs spectrum and
  oblique corrections}},
  \href{http://dx.doi.org/10.1103/PhysRevD.64.053004}{\emph{Phys. Rev. D}
  \textbf{ 64} (2001) 053004},
  [\href{https://arxiv.org/abs/hep-ph/0102144}{\texttt{hep-ph/0102144}}].

\bibitem{Grimus:2008nb}
W.~Grimus, L.~Lavoura, O.~M. Ogreid and P.~Osland, \textit{{The Oblique
  parameters in multi-Higgs-doublet models}},
  \href{http://dx.doi.org/10.1016/j.nuclphysb.2008.04.019}{\emph{Nucl. Phys. B}
  \textbf{ 801} (2008) 81--96},
  [\href{https://arxiv.org/abs/0802.4353}{\texttt{0802.4353}}].

\bibitem{Zyla:2020zbs}
{\scshape Particle Data Group} collaboration, P.~A. Zyla et~al.,
  \textit{{Review of Particle Physics}},
  \href{http://dx.doi.org/10.1093/ptep/ptaa104}{\emph{PTEP} \textbf{ 2020}
  (2020) 083C01}.

\bibitem{ParticleDataGroup:2022pth}
{\scshape Particle Data Group} collaboration, R.~L. Workman et~al.,
  \textit{{Review of Particle Physics}},
  \href{http://dx.doi.org/10.1093/ptep/ptac097}{\emph{PTEP} \textbf{ 2022}
  (2022) 083C01}.

\bibitem{Arbey:2017gmh}
A.~Arbey, F.~Mahmoudi, O.~Stal and T.~Stefaniak, \textit{{Status of the Charged
  Higgs Boson in Two Higgs Doublet Models}},
  \href{http://dx.doi.org/10.1140/epjc/s10052-018-5651-1}{\emph{Eur. Phys. J.
  C} \textbf{ 78} (2018) 182},
  [\href{https://arxiv.org/abs/1706.07414}{\texttt{1706.07414}}].

\bibitem{Sanyal:2019xcp}
P.~Sanyal, \textit{{Limits on the Charged Higgs Parameters in the Two Higgs
  Doublet Model using CMS $\sqrt{s}=13$ TeV Results}},
  \href{http://dx.doi.org/10.1140/epjc/s10052-019-7431-y}{\emph{Eur. Phys. J.
  C} \textbf{ 79} (2019) 913},
  [\href{https://arxiv.org/abs/1906.02520}{\texttt{1906.02520}}].

\bibitem{Misiak:2017bgg}
M.~Misiak and M.~Steinhauser, \textit{{Weak radiative decays of the B meson and
  bounds on $M_{H^\pm }$ in the Two-Higgs-Doublet Model}},
  \href{http://dx.doi.org/10.1140/epjc/s10052-017-4776-y}{\emph{Eur. Phys. J.
  C} \textbf{ 77} (2017) 201},
  [\href{https://arxiv.org/abs/1702.04571}{\texttt{1702.04571}}].

\bibitem{Bechtle:2020uwn}
P.~Bechtle, S.~Heinemeyer, T.~Klingl, T.~Stefaniak, G.~Weiglein and
  J.~Wittbrodt, \textit{{HiggsSignals-2: Probing new physics with precision
  Higgs measurements in the LHC 13 TeV era}},
  \href{http://dx.doi.org/10.1140/epjc/s10052-021-08942-y}{\emph{Eur. Phys. J.
  C} \textbf{ 81} (2021) 145},
  [\href{https://arxiv.org/abs/2012.09197}{\texttt{2012.09197}}].

\bibitem{Aaboud:2018gay}
{\scshape ATLAS} collaboration, M.~Aaboud et~al., \textit{{Search for Higgs
  bosons produced via vector-boson fusion and decaying into bottom quark pairs
  in $\sqrt{s} = 13$ $\mathrm{TeV}$ $pp$ collisions with the ATLAS detector}},
  \href{http://dx.doi.org/10.1103/PhysRevD.98.052003}{\emph{Phys. Rev. D}
  \textbf{ 98} (2018) 052003},
  [\href{https://arxiv.org/abs/1807.08639}{\texttt{1807.08639}}].

\bibitem{Aaboud:2018jqu}
{\scshape ATLAS} collaboration, M.~Aaboud et~al., \textit{{Measurements of
  gluon-gluon fusion and vector-boson fusion Higgs boson production
  cross-sections in the $H \to WW^{\ast} \to e\nu\mu\nu$ decay channel in $pp$
  collisions at $\sqrt{s}=13$ TeV with the ATLAS detector}},
  \href{http://dx.doi.org/10.1016/j.physletb.2018.11.064}{\emph{Phys. Lett. B}
  \textbf{ 789} (2019) 508--529},
  [\href{https://arxiv.org/abs/1808.09054}{\texttt{1808.09054}}].

\bibitem{Aaboud:2018pen}
{\scshape ATLAS} collaboration, M.~Aaboud et~al., \textit{{Cross-section
  measurements of the Higgs boson decaying into a pair of $\tau$-leptons in
  proton-proton collisions at $\sqrt{s}=13$ TeV with the ATLAS detector}},
  \href{http://dx.doi.org/10.1103/PhysRevD.99.072001}{\emph{Phys. Rev. D}
  \textbf{ 99} (2019) 072001},
  [\href{https://arxiv.org/abs/1811.08856}{\texttt{1811.08856}}].

\bibitem{Aad:2020mkp}
{\scshape ATLAS} collaboration, G.~Aad et~al., \textit{{Higgs boson production
  cross-section measurements and their EFT interpretation in the $4\ell $ decay
  channel at $\sqrt{s}=$13 TeV with the ATLAS detector}},
  \href{http://dx.doi.org/10.1140/epjc/s10052-020-8227-9}{\emph{Eur. Phys. J.
  C} \textbf{ 80} (2020) 957},
  [\href{https://arxiv.org/abs/2004.03447}{\texttt{2004.03447}}].

\bibitem{Sirunyan:2018mvw}
{\scshape CMS} collaboration, A.~M. Sirunyan et~al., \textit{{Search for $
  \mathrm{t}\overline{\mathrm{t}}\mathrm{H} $ production in the $ \mathrm{H}\to
  \mathrm{b}\overline{\mathrm{b}} $ decay channel with leptonic $
  \mathrm{t}\overline{\mathrm{t}} $ decays in proton-proton collisions at $
  \sqrt{s}=13 $ TeV}},
  \href{http://dx.doi.org/10.1007/JHEP03(2019)026}{\emph{JHEP} \textbf{ 03}
  (2019) 026}, [\href{https://arxiv.org/abs/1804.03682}{\texttt{1804.03682}}].

\bibitem{Sirunyan:2018hbu}
{\scshape CMS} collaboration, A.~M. Sirunyan et~al., \textit{{Search for the
  Higgs boson decaying to two muons in proton-proton collisions at $\sqrt{s} =$
  13 TeV}}, \href{http://dx.doi.org/10.1103/PhysRevLett.122.021801}{\emph{Phys.
  Rev. Lett.} \textbf{ 122} (2019) 021801},
  [\href{https://arxiv.org/abs/1807.06325}{\texttt{1807.06325}}].

\bibitem{CMS:2019chr}
{\scshape CMS} collaboration, \textit{{Measurements of properties of the Higgs
  boson in the four-lepton final state in proton-proton collisions at
  $\sqrt{s}=13~\mathrm{TeV}$}},  [\href{http://cds.cern.ch/record/2668684}{\texttt{http://cds.cern.ch/record/2668684}}].

\bibitem{CMS:2019kqw}
{\scshape CMS} collaboration, \textit{{Measurements of differential Higgs boson
  production cross sections in the leptonic WW decay mode at $\sqrt{s} =
  13~\mathrm{TeV}$}}, [\href{http://cds.cern.ch/record/2691268}{\texttt{http://cds.cern.ch/record/2691268}}].

\bibitem{Bechtle:2020pkv}
P.~Bechtle, D.~Dercks, S.~Heinemeyer, T.~Klingl, T.~Stefaniak, G.~Weiglein
  et~al., \textit{{HiggsBounds-5: Testing Higgs Sectors in the LHC 13 TeV
  Era}}, \href{http://dx.doi.org/10.1140/epjc/s10052-020-08557-9}{\emph{Eur.
  Phys. J. C} \textbf{ 80} (2020) 1211},
  [\href{https://arxiv.org/abs/2006.06007}{\texttt{2006.06007}}].

\bibitem{CDF:2022hxs}
{\scshape CDF} collaboration, T.~Aaltonen et~al., \textit{{High-precision
  measurement of the W boson mass with the CDF II detector}},
  \href{http://dx.doi.org/10.1126/science.abk1781}{\emph{Science} \textbf{ 376}
  (2022) 170--176}.

\bibitem{Lu:2022bgw}
C.-T. Lu, L.~Wu, Y.~Wu and B.~Zhu, \textit{{Electroweak precision fit and new
  physics in light of the W boson mass}},
  \href{http://dx.doi.org/10.1103/PhysRevD.106.035034}{\emph{Phys. Rev. D}
  \textbf{ 106} (2022) 035034},
  [\href{https://arxiv.org/abs/2204.03796}{\texttt{2204.03796}}].

\bibitem{Fan:2022dck}
Y.-Z. Fan, T.-P. Tang, Y.-L.~S. Tsai and L.~Wu, \textit{{Inert Higgs Dark
  Matter for CDF II W-Boson Mass and Detection Prospects}},
  \href{http://dx.doi.org/10.1103/PhysRevLett.129.091802}{\emph{Phys. Rev.
  Lett.} \textbf{ 129} (2022) 091802},
  [\href{https://arxiv.org/abs/2204.03693}{\texttt{2204.03693}}].

\bibitem{Zhu:2022tpr}
C.-R. Zhu, M.-Y. Cui, Z.-Q. Xia, Z.-H. Yu, X.~Huang, Q.~Yuan et~al.,
  \textit{{Explaining the GeV antiproton/$\gamma-$ray excesses and W-boson mass
  anomaly in an inert two Higgs doublet model}},
  \href{https://arxiv.org/abs/2204.03767}{\texttt{2204.03767}}.

\bibitem{Zhu:2022scj}
B.-Y. Zhu, S.~Li, J.-G. Cheng, R.-L. Li and Y.-F. Liang, \textit{{Using
  gamma-ray observation of dwarf spheroidal galaxy to test a dark matter model
  that can interpret the W-boson mass anomaly}},
  \href{https://arxiv.org/abs/2204.04688}{\texttt{2204.04688}}.

\bibitem{Song:2022xts}
H.~Song, W.~Su and M.~Zhang, \textit{{Electroweak phase transition in 2HDM
  under Higgs, Z-pole, and W precision measurements}},
  \href{http://dx.doi.org/10.1007/JHEP10(2022)048}{\emph{JHEP} \textbf{ 10}
  (2022) 048}, [\href{https://arxiv.org/abs/2204.05085}{\texttt{2204.05085}}].

\bibitem{Bahl:2022xzi}
H.~Bahl, J.~Braathen and G.~Weiglein, \textit{{New physics effects on the
  W-boson mass from a doublet extension of the SM Higgs sector}},
  \href{http://dx.doi.org/10.1016/j.physletb.2022.137295}{\emph{Phys. Lett. B}
  \textbf{ 833} (2022) 137295},
  [\href{https://arxiv.org/abs/2204.05269}{\texttt{2204.05269}}].

\bibitem{Heo:2022dey}
Y.~Heo, D.-W. Jung and J.~S. Lee, \textit{{Impact of the CDF W-mass anomaly on
  two Higgs doublet model}},
  \href{http://dx.doi.org/10.1016/j.physletb.2022.137274}{\emph{Phys. Lett. B}
  \textbf{ 833} (2022) 137274},
  [\href{https://arxiv.org/abs/2204.05728}{\texttt{2204.05728}}].

\bibitem{Babu:2022pdn}
K.~S. Babu, S.~Jana and V.~P. K., \textit{{Correlating W-Boson Mass Shift with
  Muon g-2 in the Two Higgs Doublet Model}},
  \href{http://dx.doi.org/10.1103/PhysRevLett.129.121803}{\emph{Phys. Rev.
  Lett.} \textbf{ 129} (2022) 121803},
  [\href{https://arxiv.org/abs/2204.05303}{\texttt{2204.05303}}].

\bibitem{Biekotter:2022abc}
T.~Biek\"otter, S.~Heinemeyer and G.~Weiglein, \textit{{Excesses in the
  low-mass Higgs-boson search and the $W$-boson mass measurement}},
  \href{https://arxiv.org/abs/2204.05975}{\texttt{2204.05975}}.

\bibitem{Ahn:2022xeq}
Y.~H. Ahn, S.~K. Kang and R.~Ramos, \textit{{Implications of New CDF-II $W$
  Boson Mass on Two Higgs Doublet Model}},
  \href{http://dx.doi.org/10.1103/PhysRevD.106.055038}{\emph{Phys. Rev. D}
  \textbf{ 106} (2022) 055038},
  [\href{https://arxiv.org/abs/2204.06485}{\texttt{2204.06485}}].

\bibitem{Han:2022juu}
X.-F. Han, F.~Wang, L.~Wang, J.~M. Yang and Y.~Zhang, \textit{{Joint
  explanation of W-mass and muon g\textendash{}2 in the 2HDM*}},
  \href{http://dx.doi.org/10.1088/1674-1137/ac7c63}{\emph{Chin. Phys. C}
  \textbf{ 46} (2022) 103105},
  [\href{https://arxiv.org/abs/2204.06505}{\texttt{2204.06505}}].

\bibitem{Arcadi:2022dmt}
G.~Arcadi and A.~Djouadi, \textit{{2HD plus light pseudoscalar model for a
  combined explanation of the possible excesses in the CDF MW measurement and
  (g-2)\ensuremath{\mu} with dark matter}},
  \href{http://dx.doi.org/10.1103/PhysRevD.106.095008}{\emph{Phys. Rev. D}
  \textbf{ 106} (2022) 095008},
  [\href{https://arxiv.org/abs/2204.08406}{\texttt{2204.08406}}].

\bibitem{Ghorbani:2022vtv}
K.~Ghorbani and P.~Ghorbani, \textit{{W-boson mass anomaly from scale invariant
  2HDM}}, \href{http://dx.doi.org/10.1016/j.nuclphysb.2022.115980}{\emph{Nucl.
  Phys. B} \textbf{ 984} (2022) 115980},
  [\href{https://arxiv.org/abs/2204.09001}{\texttt{2204.09001}}].

\bibitem{Broggio:2014mna}
A.~Broggio, E.~J. Chun, M.~Passera, K.~M. Patel and S.~K. Vempati,
  \textit{{Limiting two-Higgs-doublet models}},
  \href{http://dx.doi.org/10.1007/JHEP11(2014)058}{\emph{JHEP} \textbf{ 11}
  (2014) 058}, [\href{https://arxiv.org/abs/1409.3199}{\texttt{1409.3199}}].

\bibitem{Oredsson:2018vio}
J.~Oredsson, \textit{{2HDME : Two-Higgs-Doublet Model Evolver}},
  \href{http://dx.doi.org/10.1016/j.cpc.2019.05.021}{\emph{Comput. Phys.
  Commun.} \textbf{ 244} (2019) 409--426},
  [\href{https://arxiv.org/abs/1811.08215}{\texttt{1811.08215}}].

\bibitem{Ivanov:2007de}
I.~P. Ivanov, \textit{{Minkowski space structure of the Higgs potential in
  2HDM. II. Minima, symmetries, and topology}},
  \href{http://dx.doi.org/10.1103/PhysRevD.77.015017}{\emph{Phys. Rev. D}
  \textbf{ 77} (2008) 015017},
  [\href{https://arxiv.org/abs/0710.3490}{\texttt{0710.3490}}].

\bibitem{Akeroyd:1995hg}
A.~G. Akeroyd, \textit{{Fermiophobic Higgs bosons at the Tevatron}},
  \href{http://dx.doi.org/10.1016/0370-2693(95)01478-0}{\emph{Phys. Lett. B}
  \textbf{ 368} (1996) 89--95},
  [\href{https://arxiv.org/abs/hep-ph/9511347}{\texttt{hep-ph/9511347}}].

\bibitem{Akeroyd:1998ui}
A.~G. Akeroyd, \textit{{Fermiophobic and other nonminimal neutral Higgs bosons
  at the LHC}}, \href{http://dx.doi.org/10.1088/0954-3899/24/11/001}{\emph{J.
  Phys. G} \textbf{ 24} (1998) 1983--1994},
  [\href{https://arxiv.org/abs/hep-ph/9803324}{\texttt{hep-ph/9803324}}].

\bibitem{Barroso:1999bf}
A.~Barroso, L.~Brucher and R.~Santos, \textit{{Is there a light fermiophobic
  Higgs?}}, \href{http://dx.doi.org/10.1103/PhysRevD.60.035005}{\emph{Phys.
  Rev. D} \textbf{ 60} (1999) 035005},
  [\href{https://arxiv.org/abs/hep-ph/9901293}{\texttt{hep-ph/9901293}}].

\bibitem{Brucher:1999tx}
L.~Brucher and R.~Santos, \textit{{Experimental signatures of fermiophobic
  Higgs bosons}}, \href{http://dx.doi.org/10.1007/s100529900252}{\emph{Eur.
  Phys. J. C} \textbf{ 12} (2000) 87--98},
  [\href{https://arxiv.org/abs/hep-ph/9907434}{\texttt{hep-ph/9907434}}].

\bibitem{Akeroyd:2003xi}
A.~G. Akeroyd, M.~A. Diaz and F.~J. Pacheco, \textit{{Double fermiophobic Higgs
  boson production at the CERN LHC and LC}},
  \href{http://dx.doi.org/10.1103/PhysRevD.70.075002}{\emph{Phys. Rev. D}
  \textbf{ 70} (2004) 075002},
  [\href{https://arxiv.org/abs/hep-ph/0312231}{\texttt{hep-ph/0312231}}].

\bibitem{Arhrib:2008pw}
A.~Arhrib, R.~Benbrik, R.~B. Guedes and R.~Santos, \textit{{Search for a light
  fermiophobic Higgs boson produced via gluon fusion at Hadron Colliders}},
  \href{http://dx.doi.org/10.1103/PhysRevD.78.075002}{\emph{Phys. Rev. D}
  \textbf{ 78} (2008) 075002},
  [\href{https://arxiv.org/abs/0805.1603}{\texttt{0805.1603}}].

\bibitem{Berger:2012sy}
E.~L. Berger, Z.~Sullivan and H.~Zhang, \textit{{Associated Higgs plus vector
  boson test of a fermiophobic Higgs boson}},
  \href{http://dx.doi.org/10.1103/PhysRevD.86.015011}{\emph{Phys. Rev. D}
  \textbf{ 86} (2012) 015011},
  [\href{https://arxiv.org/abs/1203.6645}{\texttt{1203.6645}}].

\bibitem{Ilisie:2014hea}
V.~Ilisie and A.~Pich, \textit{{Low-mass fermiophobic charged Higgs
  phenomenology in two-Higgs-doublet models}},
  \href{http://dx.doi.org/10.1007/JHEP09(2014)089}{\emph{JHEP} \textbf{ 09}
  (2014) 089}, [\href{https://arxiv.org/abs/1405.6639}{\texttt{1405.6639}}].

\bibitem{Delgado:2016arn}
A.~Delgado, M.~Garcia-Pepin, M.~Quiros, J.~Santiago and R.~Vega-Morales,
  \textit{{Diphoton and Diboson Probes of Fermiophobic Higgs Bosons at the
  LHC}}, \href{http://dx.doi.org/10.1007/JHEP06(2016)042}{\emph{JHEP} \textbf{
  06} (2016) 042},
  [\href{https://arxiv.org/abs/1603.00962}{\texttt{1603.00962}}].

\bibitem{Mondal:2021bxa}
T.~Mondal and P.~Sanyal, \textit{{Same sign trilepton as signature of charged
  Higgs in two Higgs doublet model}},
  \href{http://dx.doi.org/10.1007/JHEP05(2022)040}{\emph{JHEP} \textbf{ 05}
  (2022) 040}, [\href{https://arxiv.org/abs/2109.05682}{\texttt{2109.05682}}].

\bibitem{Cheung:2022ndq}
K.~Cheung, A.~Jueid, J.~Kim, S.~Lee, C.-T. Lu and J.~Song,
  \textit{{Comprehensive study of the light charged Higgs boson in the type-I
  two-Higgs-doublet model}},
  \href{http://dx.doi.org/10.1103/PhysRevD.105.095044}{\emph{Phys. Rev. D}
  \textbf{ 105} (2022) 095044},
  [\href{https://arxiv.org/abs/2201.06890}{\texttt{2201.06890}}].

\bibitem{Forslund:2022xjq}
M.~Forslund and P.~Meade, \textit{{High precision higgs from high energy muon
  colliders}}, \href{http://dx.doi.org/10.1007/JHEP08(2022)185}{\emph{JHEP}
  \textbf{ 08} (2022) 185},
  [\href{https://arxiv.org/abs/2203.09425}{\texttt{2203.09425}}].
  
\bibitem{Misiak:2020vlo}
M.~Misiak, A.~Rehman and M.~Steinhauser,
\textit{Towards $ \overline{B}\to {X}_s\gamma $ at the NNLO in QCD without interpolation in m$_{c}$,} 
 \href{http://dx.doi.org/10.1007/JHEP06(2020)175}{\emph{JHEP} \textbf{ 06} (2020) 175},
  [\href{https://arxiv.org/abs/2002.01548}{\texttt{2203.09425}}].

\bibitem{Djouadi:2005gj}
A.~Djouadi, \textit{{The Anatomy of electro-weak symmetry breaking. II. The
  Higgs bosons in the minimal supersymmetric model}},
  \href{http://dx.doi.org/10.1016/j.physrep.2007.10.005}{\emph{Phys. Rept.}
  \textbf{ 459} (2008) 1--241},
  [\href{https://arxiv.org/abs/hep-ph/0503173}{\texttt{hep-ph/0503173}}].

\bibitem{Degrande:2011ua}
C.~Degrande, C.~Duhr, B.~Fuks, D.~Grellscheid, O.~Mattelaer and T.~Reiter,
  \textit{{UFO - The Universal FeynRules Output}},
  \href{http://dx.doi.org/10.1016/j.cpc.2012.01.022}{\emph{Comput. Phys.
  Commun.} \textbf{ 183} (2012) 1201--1214},
  [\href{https://arxiv.org/abs/1108.2040}{\texttt{1108.2040}}].

\bibitem{Alloul:2013bka}
A.~Alloul, N.~D. Christensen, C.~Degrande, C.~Duhr and B.~Fuks,
  \textit{{FeynRules 2.0 - A complete toolbox for tree-level phenomenology}},
  \href{http://dx.doi.org/10.1016/j.cpc.2014.04.012}{\emph{Comput. Phys.
  Commun.} \textbf{ 185} (2014) 2250--2300},
  [\href{https://arxiv.org/abs/1310.1921}{\texttt{1310.1921}}].

\bibitem{Alwall:2011uj}
J.~Alwall, M.~Herquet, F.~Maltoni, O.~Mattelaer and T.~Stelzer,
  \textit{{MadGraph 5 : Going Beyond}},
  \href{http://dx.doi.org/10.1007/JHEP06(2011)128}{\emph{JHEP} \textbf{ 06}
  (2011) 128}, [\href{https://arxiv.org/abs/1106.0522}{\texttt{1106.0522}}].

\bibitem{NNPDF:2017mvq}
{\scshape NNPDF} collaboration, R.~D. Ball et~al., \textit{{Parton
  distributions from high-precision collider data}},
  \href{http://dx.doi.org/10.1140/epjc/s10052-017-5199-5}{\emph{Eur. Phys. J.
  C} \textbf{ 77} (2017) 663},
  [\href{https://arxiv.org/abs/1706.00428}{\texttt{1706.00428}}].

\bibitem{Baglio:2012np}
J.~Baglio, A.~Djouadi, R.~Gr\"ober, M.~M. M\"uhlleitner, J.~Quevillon and
  M.~Spira, \textit{{The measurement of the Higgs self-coupling at the LHC:
  theoretical status}},
  \href{http://dx.doi.org/10.1007/JHEP04(2013)151}{\emph{JHEP} \textbf{ 04}
  (2013) 151}, [\href{https://arxiv.org/abs/1212.5581}{\texttt{1212.5581}}].

\bibitem{ATLAS:2018rvj}
{\scshape ATLAS} collaboration, \textit{{Measurement prospects of the pair
  production and self-coupling of the Higgs boson with the ATLAS experiment at
  the HL-LHC}}, [\href{http://cds.cern.ch/record/2652727}{\texttt{http://cds.cern.ch/record/2652727}}].


\bibitem{Ballestrero:2008gf}
A.~Ballestrero, G.~Bevilacqua and E.~Maina, \textit{{A Complete parton level
  analysis of boson-boson scattering and ElectroWeak Symmetry Breaking in lv +
  four jets production at the LHC}},
  \href{http://dx.doi.org/10.1088/1126-6708/2009/05/015}{\emph{JHEP} \textbf{
  05} (2009) 015},
  [\href{https://arxiv.org/abs/0812.5084}{\texttt{0812.5084}}].

\bibitem{Jung:2021tym}
S.~Jung, Z.~Liu, L.-T. Wang and K.-P. Xie, \textit{{Probing Higgs boson exotic
  decays at the LHC with machine learning}},
  \href{http://dx.doi.org/10.1103/PhysRevD.105.035008}{\emph{Phys. Rev. D}
  \textbf{ 105} (2022) 035008},
  [\href{https://arxiv.org/abs/2109.03294}{\texttt{2109.03294}}].

\bibitem{Gomez-Ceballos:2013zzn}
{\scshape TLEP Design Study Working Group} collaboration, M.~Bicer et~al.,
  \textit{{First Look at the Physics Case of TLEP}},
  \href{http://dx.doi.org/10.1007/JHEP01(2014)164}{\emph{JHEP} \textbf{ 01}
  (2014) 164}, [\href{https://arxiv.org/abs/1308.6176}{\texttt{1308.6176}}].

\bibitem{Gao:2021bam}
J.~Gao, \textit{{CEPC and SppC Status \textemdash{} From the completion of CDR
  towards TDR}}, \href{http://dx.doi.org/10.1142/S0217751X21420057}{\emph{Int.
  J. Mod. Phys. A} \textbf{ 36} (2021) 2142005}.

\bibitem{CEPCStudyGroup:2018ghi}
{\scshape CEPC Study Group} collaboration, M.~Dong et~al., \textit{{CEPC
  Conceptual Design Report: Volume 2 - Physics \& Detector}},
  \href{https://arxiv.org/abs/1811.10545}{\texttt{1811.10545}}.

\bibitem{Palmer:2007zzc}
R.~B. Palmer, J.~S. Berg, R.~C. Fernow, J.~C. Gallardo, H.~G. Kirk, Y.~Alexahin
  et~al., \textit{{A Complete Scheme of Ionization Cooling for a Muon
  Collider}}, \href{http://dx.doi.org/10.2172/921988}{\emph{Conf. Proc. C}
  \textbf{ 070625} (2007) 3193},
  [\href{https://arxiv.org/abs/0711.4275}{\texttt{0711.4275}}].

\bibitem{Delahaye:2019omf}
J.~P. Delahaye, M.~Diemoz, K.~Long, B.~Mansouli\'e, N.~Pastrone, L.~Rivkin
  et~al., \textit{{Muon Colliders}},
  \href{https://arxiv.org/abs/1901.06150}{\texttt{1901.06150}}.

\bibitem{Black:2022cth}
K.~M. Black et~al., \textit{{Muon Collider Forum Report}},
  \href{https://arxiv.org/abs/2209.01318}{\texttt{2209.01318}}.

\bibitem{Cheng:1973nv}
T.~P. Cheng, E.~Eichten and L.-F. Li, \textit{{Higgs Phenomena in
  Asymptotically Free Gauge Theories}},
  \href{http://dx.doi.org/10.1103/PhysRevD.9.2259}{\emph{Phys. Rev. D} \textbf{
  9} (1974) 2259}.

\bibitem{Komatsu:1981xh}
H.~Komatsu, \textit{{Behavior of the Yukawa and the Quartic Scalar Couplings in
  Grand Unified Theories}},
  \href{http://dx.doi.org/10.1143/PTP.67.1177}{\emph{Prog. Theor. Phys.}
  \textbf{ 67} (1982) 1177}.

\end{thebibliography}

\end{document}